\documentclass[a4paper,11pt]{article}
\pdfoutput=1
\usepackage{jheppub}

\usepackage{tikz}

\usepackage{enumitem}
\usepackage[T1]{fontenc} 
\usepackage{floatrow}
\usepackage{appendix}
\usepackage{tabularx}
\usepackage[normalem]{ulem}
\usepackage{multicol}
\usepackage{mwe}

\allowdisplaybreaks

\usepackage{epsf}
\usepackage{amsmath}
\usepackage{amsfonts}
\usepackage{amssymb}
\usepackage{psfrag,epsfig,graphicx,graphics}

\newcommand\numberthis[1][]{%
    \refstepcounter{equation}%
    \ifx#1\empty\else\label{eq:#1}\fi%
    \tag{\theequation}%
}

\usepackage{xargs} 
\usepackage[colorinlistoftodos,prependcaption,textsize=tiny]{todonotes}
\newcommandx{\APa}[2][1=]{\todo[linecolor=blue,backgroundcolor=blue!25,bordercolor=blue,#1]{#2}}

\newcommandx{\APo}[2][1=]{\todo[linecolor=black,backgroundcolor=black!25,bordercolor=black,#1]{#2}}

\providecommand{\U}[1]{\protect\rule{.1in}{.1in}}

\def\slashchar#1{\setbox0=\hbox{$#1$}
   \dimen0=\wd0
   \setbox1=\hbox{/} \dimen1=\wd1
   \ifdim\dimen0>\dimen1
      \rlap{\hbox to \dimen0{\hfil/\hfil}}
      #1
   \else
      \rlap{\hbox to \dimen1{\hfil$#1$\hfil}}
      /
   \fi}

\def\bei{\begin{itemize}}
\def\ei{\end{itemize}}

\def\beeq{\begin{eqnarray}} 
\def\beqa{\begin{eqnarray}}
\def\bea{\begin{eqnarray}}

\def\eea{\end{eqnarray}}
\def\eqa{\end{eqnarray}}
\def\eeeq{\end{eqnarray}}

\def\eqar{\end{array}}
\def\beqar{\begin{array}}

\def\beas{\begin{eqnarray*}}
\def\beqas{\begin{eqnarray*}}

\def\eqas{\end{eqnarray*}}
\def\eeas{\end{eqnarray*}}

\def\beq{\begin{equation}} 
\def\be{\begin{equation}}

\def\ee{\end{equation}}
\def\eq{\end{equation}}
\def\eeq{\end{equation}}

\def\beqd{\begin{displaymath}}
\def\eeqd{\end{displaymath}}
\def\eqd{\end{displaymath}}

\def\beeq{\begin{eqnarray}} \def\eeeq{\end{eqnarray}}

\newcommand{\fin}{\end{document}}

\title{\boldmath High-energy factorization via eigenfunctions of the next-to-leading-order BFKL kernel}

\author[a]{Ada Polizzi,\footnote{Corresponding author}}
\author[b]{Michael Fucilla,}
\author[a,c]{Alessandro Papa}

\affiliation[a]{Dipartimento di Fisica, Università della Calabria, Arcavacata di Rende, I-87036, Cosenza, Italy}

\affiliation[b]{National Centre for Nuclear Research, Pasteura 7, Warsaw 02-093, Poland}

\affiliation[c]{INFN, Gruppo Collegato di Cosenza, Arcavacata di Rende, I-87036, Cosenza, Italy}

\emailAdd{polizziada2@gmail.com}
\emailAdd{michael.fucilla@ijclab.in2p3.fr}
\emailAdd{alessandro.papa@fis.unical.it}

\abstract{We present a general formula for the amplitude of forward exclusive hadronic processes in the semihard regime of perturbative Quantum Chromodynamics (QCD), by means of the {\em next-to-leading order} eigenfunctions of the Balitsky-Fadin-Kuraev-Lipatov (BFKL) kernel, as constructed by Chirilli and Kovchegov. We discuss some formal subtleties in the check of compatibility with the similar formula based on the use of the {\em leading-order} BFKL eigenfunctions. Finally, in the specific case of the electroproduction of two light vector mesons, we consider the numerical stability of the amplitude when one or the other set of eigenfunctions is adopted.}

\begin{document} 
\maketitle
\flushbottom

\section{Introduction}
\label{sec:intro}

Perturbative Quantum Chromodynamics (QCD) is the reference theory for the study of hadronic processes characterized by the presence of an energy scale $Q$ (a {\it hard} scale) much larger than the scale of hadronic masses, given by $\Lambda_{\rm QCD}$. At the LHC, however, unprecedentedly large center-of-mass energies $\sqrt{s}$ have been reached, so to explore the kinematic region (called {\it semihard} region) where  
\[
s \gg Q^2 \gg \Lambda^2_{\rm QCD} \;.
\]
In this regime, fixed-order QCD perturbation theory becomes inefficient, due to the appearance in the coefficients of the perturbative series of large logarithms of $s$ which compensate the smallness of the QCD coupling $\alpha_s$, thus calling for an
all-order resummation.

The Balitsky-Fadin-Kuraev-Lipatov (BFKL) approach~\cite{Fadin:1975cb,Kuraev:1976ge,Kuraev:1977fs,Balitsky:1978ic} 
has long been established as the theoretical framework for the study of semihard processes, whereby a procedure has been developed for the resummation of leading energy logs, {\it i.e.} all terms proportional to $[\alpha_s\ln s]^n$ (the {\it leading-logarithmic approximation} or LLA), and of next-to-leading ones, {\it i.e.} all terms of the form $\alpha_s[\alpha_s\ln s]^n$ (the next-to-LLA approximation or NLLA).
Within this approach a neat separation is achieved between the longitudinal dynamics,
which is independent of the considered semihard process and is encoded in a universal {\it Green function}, and the transverse one, which manifests in two process-dependent {\it impact factors}, describing the transition of each colliding particle to a definite state in its fragmentation region. The bottom-line is that the $s$-channel imaginary part of an amplitude takes a peculiarly factorized form, in the transverse momentum space, of the impact factors and the Green function. 

The BFKL Green function is defined by an iterative integral equation and its form is determined by the {\it kernel} entering this equation: in the LLA it is enough to know the kernel at the leading order (LO), while the next-to-LO (NLO) kernel is needed to accomplish resummation with NLLA. The NLO BFKL kernel is available both for forward scattering ({\it i.e.} for $t = 0$ and color singlet in the $t$-channel)~\cite{Fadin:1998py,Ciafaloni:1998gs} and for any fixed, not growing with $s$, momentum transfer $t$ and any possible two-gluon colored exchange in the $t$-channel~\cite{Fadin:1998jv,Fadin:2000kx,Fadin:2000hu,Fadin:2004zq,Fadin:2005zj}.
Some pieces of the next-to-NLO kernel have been recently found in $\mathcal{N}=4$ SYM~\cite{Byrne:2022wzk}, in pure-gauge QCD~\cite{DelDuca:2021vjq} and in full QCD~\cite{Caola:2021izf,Falcioni:2021dgr,Fadin:2023roz,Abreu:2024mpk,Buccioni:2024gzo,Abreu:2024xoh}.

A full NLLA implementation requires that impact factors are calculated in the NLO, which is usually tough. The first to appear were the quark and gluon impact factors~\cite{Fadin:1999de,Fadin:1999df,Ciafaloni:1998kx,Ciafaloni:1998hu,Ciafaloni:2000sq}, which are closely related to the forward-jet ones~\cite{Bartels:2001ge,Bartels:2002yj,Caporale:2011cc,Ivanov:2012ms,Colferai:2015zfa,Chirilli:2012jd, Hentschinski:2014bra}. Subsequently, within distinct small-$x$ formalism, a prolific research activity led to several full NLO calculations, comprising inclusive~\cite{Bartels:2000gt,Bartels:2001mv,Bartels:2002uz,Bartels:2003zi,Bartels:2004bi,Fadin:2001ap,Balitsky:2012bs,Beuf:2022ndu} and diffractive~\cite{Beuf:2022kyp} deep-inelastic scattering (DIS), photon-dijet production in DIS~\cite{Roy:2019hwr}, dijets production in inclusive~\cite{Caucal:2021ent,Caucal:2022ulg} and diffractive~\cite{Boussarie:2014lxa,Boussarie:2016ogo} DIS, dijet photoproduction~\cite{Taels:2022tza}, single and dihadron production in both inclusive and diffractive DIS~\cite{Bergabo:2022zhe,Bergabo:2022tcu,Iancu:2022gpw,Caucal:2024nsb,Fucilla:2023mkl,Fucilla:2022wcg}, exclusive light vector meson production~\cite{Ivanov:2012iv,Boussarie:2016bkq,Mantysaari:2022bsp}, inclusive and exclusive quarkonium production~\cite{Nefedov:2023uen,Nefedov:2024swu,Mantysaari:2022kdm}, forward Higgs~\cite{Nefedov:2019mrg,Hentschinski:2020tbi,Celiberto:2022fgx,Fucilla:2024cpf,Celiberto:2024bbv}~\footnote{Recently, in the infinite top-mass approximation, the two-loop corrections to this impact factor were obtained~\cite{DelDuca:2025vux}.}, forward Drell-Yan~\cite{Taels:2023czt}.

As we will explicitly see in the next Section, two mathematical tools are particularly useful in working out all the ingredients of the BFKL approach and in extracting physical predictions, both analytically and numerically: the {\it Mellin transform} with respect to the variable $s$, leading to the appearance of its conjugate variable which we will denote by $\omega$, and the {\it eigenfunctions of the kernel}, which arise considering the BFKL kernel as an operator acting in the transverse momentum space. The Mellin transform is useful, first of all, because it allows to write the 
iterative equation defining the BFKL Green function $G$, {\it i.e.} the BFKL equation, in a simple fashion and, more importantly, because the position of singularities of the Mellin transform of the Green function ($G_\omega$) in the $\omega$-complex plane determines the (universal) leading behavior in $s$ of amplitudes. The knowledge of the kernel eigenfunctions allows to determine
the leading singularities of $G_\omega$ and to disentangle the convolution of Green function and impact factors in the transverse momentum space. Moreover, rewriting amplitudes in the BFKL approach in terms of the eigenfunctions of the kernel
facilitates their numerical implementation. Indeed, many phenomenological applications of the BFKL approach so far made use of representations of the BFKL amplitude in terms of the eigenfunctions of the kernel (see Ref.~\cite{Celiberto:2020wpk} for a review, and
Refs.~\cite{Celiberto:2022dyf,Celiberto:2024beg,Celiberto:2024mab} for some recent applications)
and, specifically, those of the LO kernel, which are known since the early days of the BFKL approach~\cite{Fadin:1975cb,Kuraev:1976ge,Kuraev:1977fs,Balitsky:1978ic}. We refer also to Ref.~\cite{Grabovsky:2013gta} for a solution of the NLO BFKL equation in terms of the LO kernel eigenfunctions. Alternative approaches to small-$x$ phenomenology rely on on Monte Carlo integration methods~\cite{Chachamis:2015zzp,Chachamis:2011rw,Andersen:2021qma,Andersen:2017sht}, which are based on a numerical iterative solution of the BFKL equation in the transverse momentum space and are therefore totally insensitive to the choice of the basis of kernel eigenfunctions. It would be interesting to compare the results of this paper with those obtained in the Monte Carlo approach.

In this paper, we employ the eigenfunctions of the NLO kernel, as constructed in Ref.~\cite{1305.1924}, to get a process-independent representation of 
the NLLA amplitude. We first show how to handle the NLO eigenfunctions, which contain terms singular in the real parameter (usually called $\nu$) labeling them, and get ready-to-use formulas, to be exploited in future phenomenological applications. Then, we check that the use of LO eigenfunctions and NLO ones leads to equivalent expressions of the amplitude within the NLLA approximation. Finally we explore, in a first preliminary application, if the use of the NLO eigenfunctions brings along some improvement in the numerical stability of the NLLA amplitude, which is known to be strongly dependent on the energy scales entering its expression (the scale $s_0$ introduced by the Mellin transform and the renormalization scale $\mu_R$).

This is the outline of the paper: in Section~\ref{sec:LO_eigen}, we recall the basics of the BFKL approach and the derivation of the form of the NLLA amplitude when written in terms of the LO kernel eigenfunctions; in Section~\ref{sec:CK}, we recall the form of the Chirilli-Kovchegov eigenfunctions of the NLO BFKL kernel; in Section~\ref{sec:NLO_eigen}, we present the NLLA amplitude when written in terms 
of the NLO kernel eigenfunctions and verify the compatibility with the representation given in Section~\ref{sec:LO_eigen};
in Section~\ref{sec:rho_prod}, we present a numerical application of the new expression of the amplitude to the case of an exclusive process, the electroproduction of two longitudinally polarized light vector mesons, and discuss the effect on the
numerical stability of results; in Section~\ref{sec:conclusions}, we draw our conclusions.

\section{BFKL amplitude via eigenfunctions of the LO BFKL kernel}
\label{sec:LO_eigen}

We concentrate on the simple case of the high-energy limit of an exclusive process $A(p_A)+B(p_B) \to A'(p'_A)+B'(p'_B)$ in the {\it forward} regime, {\it i.e.} when the transverse momenta of produced particles are zero. This includes as a special case  the {\it elastic} process $A(p_A)+B(p_B) \to A(p_A)+B(p_B)$ at zero transferred momentum, whose $s$-channel imaginary part is connected by the optical theorem to the total cross section $A(p_A)+B(p_B) \to$ all. For the purposes of this work, it is irrelevant if $A, A'$ and $B, B'$ represent hadronic or partonic states. 
The extension to high-energy inelastic processes, with identified particles or jets in the forward and backward regions and undetected radiation in the central rapidity region (the so-called {\em forward-backward} semihard processes), is straightforward, as far as the content of the present Section is concerned.

In the limit $s=(p_A+p_B)^2\to \infty$, the BFKL approach allows to obtain a factorized form for the amplitude, characterized by a clear separation between longitudinal and transverse momentum variables, as defined by the
Sudakov decomposition 
\begin{equation}
p = \beta p_1 + \alpha p_2 + p_{\perp}\;,
\label{Sudakov1}
\end{equation}
the vectors $p_1,\ p_2$ being the light-cone basis of the initial
particle momenta plane $p_A,\ p_B$, so that
\begin{equation}
p_A= p_1 +
\frac{m_A^2}{s}p_2~,\ \ \ \ \ \ \ p_B = p_2 +
\frac{m_B^2}{s} p_1~.
\label{Sudakov2}
\end{equation}
Here $m_A$ and $m_B$ are the masses of the colliding particles $A$ and $B$, respectively; moreover, the transverse vector $p_\perp$ is always space-like, which justifies the notation $p_{\perp}^2 = - \vec p^{\:2}$.

\begin{figure}
\includegraphics[scale=0.5]{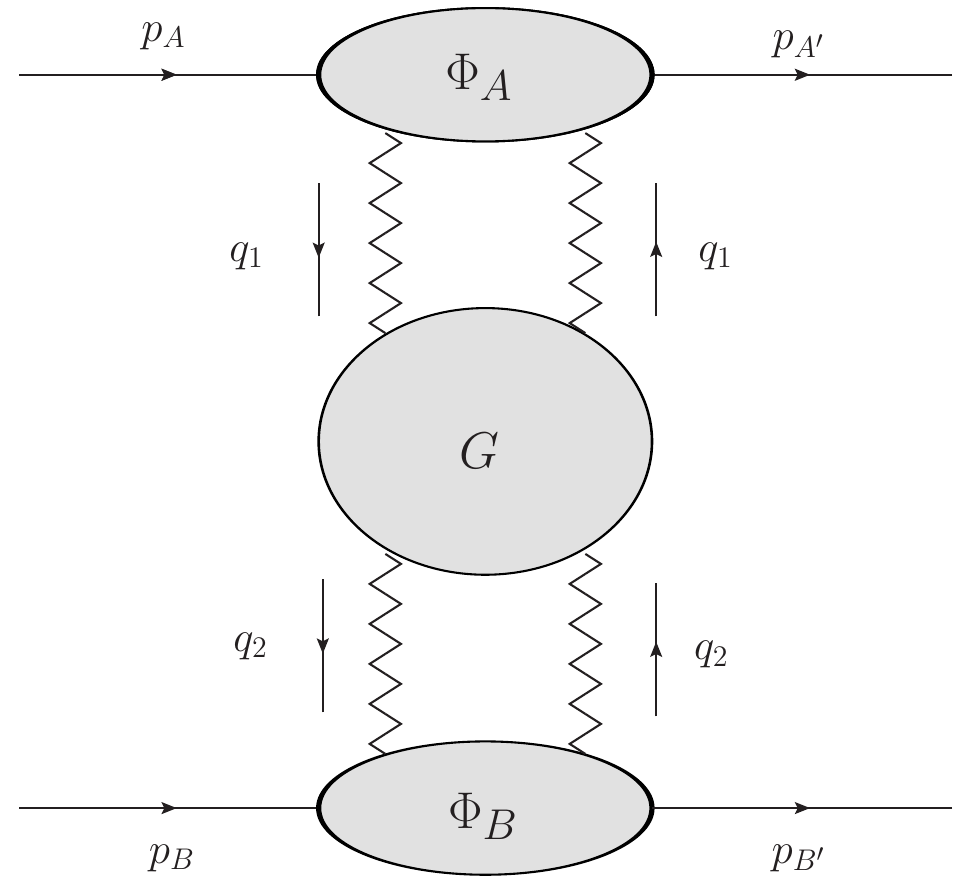}
\caption{Schematic representation of the BFKL amplitude for a forward process.}
\label{fig:BFKL_ampl}
\end{figure}

Within the NLLA, the elastic scattering amplitude for $t=0$ can be written as
(see Fig.~\ref{fig:BFKL_ampl})
\[
\Im m_{s}{\cal A} = \frac{s}{\left( 2\pi \right)^2}
\int \frac{d^2 q_1}{\vec{q}_{1}^{\:2}}
\int \frac{d^2 q_2}{\vec{q}_{2}^{\:2}}
\]
\begin{equation}
\times \Phi_{A}\left( \vec{q}_{1};s_{0}\right)\int_{\delta -i\infty}^{\delta+i\infty}
\frac{d\omega }{2\pi i}\left[ \left( \frac{s}{s_{0}}\right)^{\omega }
G_{\omega }\left( \vec{q}_{1},\vec{q}_{2}\right) 
\right] \Phi_{B}\left( -\vec{q}_{2};s_{0}\right) \;.
\label{Ar}
\end{equation} 
Here, $\Phi_{A,B}$ are the so-called impact factors and $G_{\omega}$ is the Mellin transform of the Green function for the Reggeon-Reggeon 
scattering~\cite{Fadin:1998fv}. 
The parameter $s_0$ is an arbitrary energy scale introduced in order to define the partial wave expansion of the scattering amplitudes and its dependence 
cancels in the amplitude, within NLLA. The line $\Re e(\omega)=\delta$ over which the integration in the complex $\omega$-plane is performed is arbitrary, provided that it lies to the right of all singularities in $\omega$ of $G_\omega$.

The Green function obeys the BFKL equation
\begin{equation}
\omega G_{\omega }\left( \vec{q}_{1},\vec{q}_{2}\right) = \delta^{(2)}\left( \vec{q}_{1}-\vec{q}_{2}\right)
+\int d^2 q_r \ K\left( \vec{q}_{1},\vec{q}_r\right) G_{\omega }\left( \vec{q}_r,\vec{q}_{2}\right) \;,
\label{genBFKL}
\end{equation}
where $K$ is the kernel in the singlet color representation; it has a 
perturbative expansion of the form
\begin{equation}
K( \vec{q}_{1},\vec{q}_2) = \bar\alpha_s \ K^{(0)}( \vec{q}_{1},\vec{q}_2)
+ \bar\alpha_s^2 \ K^{(1)}( \vec{q}_{1},\vec{q}_2)\;, \;\;\;\;\;
\bar\alpha_s \equiv \frac{N_c}{\pi}\alpha_s\;,
\label{K_expansion}
\end{equation}
$N_c$ being the number of colors. The expression of  $K^{(0)}$ is given in the Appendix, while the explicit form of $K^{(1)}$, which can be found, {\it e.g.}, in~\cite{Fadin:1998sh}, is not relevant here. The general definition of impact factors can be found in Ref.~\cite{Fadin:1998fv}; also for them we can write an expansion of the form
\begin{eqnarray}
\Phi_{A}(\vec{q}_{1};s_{0}) &=& \alpha_s \Biggl[\Phi_A^{(0)}(\vec{q}_{1};s_0) + \bar\alpha_s \ \Phi_A^{(1)}( \vec{q}_{1};s_0)\Biggr]\;, \nonumber \\
\Phi_{B}(-\vec{q}_{2};s_{0}) &=& \alpha_s \Biggl[ \Phi_B^{(0)}(-\vec{q}_{2};s_0) + \bar\alpha_s \ \Phi_B^{(1)}( -\vec{q}_{2};s_0)\Biggr]\;.\label{phi_expansion}
\end{eqnarray}
It is convenient to introduce a notation in which impact factors are understood as ``states'' in the space of transverse momenta, while kernel and Green function act as ``operators'':
\[
\hat{\vec q} \ |\vec q_i\rangle = \vec q_i \  |\vec q_i\rangle\;, \;\;\;\;\;
\langle \vec q_1 | \vec q_2 \rangle = \delta^{(2)}(\vec q_1 - \vec q_2)\;,
\]
\[
\langle F | G \rangle = \int d^2k \, \langle F|\vec k \rangle \langle \vec k | G \rangle = \int d^2k \, F(\vec k) G(\vec k)\;,
\]
\[
\langle \vec q_1 | \hat K | \vec q_2 \rangle = K(\vec q_1, -\vec q_2)\;,
\]
where the last equation actually defines the function $K$ and the minus sign in the second argument reflects the fact that $\vec q_2$ is "outgoing", while $\vec q_1$ is ingoing (see Fig.~\ref{fig:BFKL_ampl}).
With this notation, the expression~(\ref{Ar}) of the amplitude can be rewritten in the following, very simple form:
\begin{equation}
\Im m_{s} {\cal A} = \frac{s}{(2\pi)^2}
\int_{\delta -i\infty}^{\delta+i\infty} \frac{d\omega }{2\pi i}
\left( \frac{s}{s_{0}}\right)^{\omega } \
\langle \Phi_{A} | \ \frac{1}{\hat{\vec q}^{\:2}}  \hat G_{\omega } 
\frac{1}{\hat{\vec q}^{\:2}} \ | \Phi_{B}\rangle \;.
\label{Ar_op}
\end{equation}
Similarly, the BFKL equation~(\ref{genBFKL}) becomes
\[
\omega \, \hat G_{\omega } = \hat 1 + \hat K \, \hat G_\omega\;,
\]
which trivially leads to the solution
\[
\hat G_{\omega } = \frac{\hat 1}{\omega - \hat K}\;.
\]
In the large-$s$ limit, the asymptotic behavior of the amplitude~(\ref{Ar_op}) is ruled by the rightmost singularity on the complex-$\omega$ plane of the integrand. Localizing this singularity amounts to solve the eigenvalue problem for the kernel operator $\hat K$. Let us recall how to work this out in the LLA case, when both kernel and impact factors must be taken at the LO. The eigenvalue problem for $\hat K^{(0)}$ was solved long ago~\cite{Fadin:1975cb,Kuraev:1976ge,Kuraev:1977fs,Balitsky:1978ic}:
\[
\hat K^{(0)}\, |\nu^{(0)},n\rangle = \chi(\nu,n) \, |\nu^{(0)},n\rangle \;, \;\;\;
\chi (\nu,n)= 2\psi(1)-\psi\left(\frac{n+1}{2}+i\nu\right)-\psi\left(\frac{n+1}{2}-i\nu\right)\;, 
\]
\beq
\langle\vec q\, |\nu^{(0)},n\rangle =\frac{1}{\pi \sqrt{2}}
\left(\vec q^{\,\, 2}\right)^{i\nu-\frac{1}{2}} e^{i n\theta}\equiv \phi^{(0)}_\nu(\vec q^{\:2}) e^{i n\theta}\;, 
\;\;\;\;\;  \frac{q_x}{|\vec q|} = \cos \theta\;,
\label{LO_eigen}
\eeq
with $\{|\nu^{(0)},n\rangle,\nu\in \mathbb{R}, n \in \mathbb{Z}\}$ forming a complete set in the transverse space, satisfying the normalization condition $\langle \nu'^{(0)},n'|\nu^{(0)},n\rangle = \delta(\nu' - \nu) \delta_{n'n}$. By suitably using the completeness relation in the $|\nu^{(0)},n\rangle$-basis, one can easily get for the LLA amplitude the following expression:
\[
\Im m_{s} {\cal A}\biggl.\biggr|_{\rm LLA} = 
\frac{s}{(2\pi)^2}
\int_{\delta -i\infty}^{\delta+i\infty} \frac{d\omega }{2\pi i}
\left( \frac{s}{s_{0}}\right)^{\omega } \
\langle \Phi_A^{(0)} \ | \ \frac{1}{\hat{\vec q}^{\:2}} \ \frac{\hat 1}{\omega - \bar\alpha_s \hat K^{(0)}} \ \frac{1}{\hat{\vec q}^{2}} \ | \Phi_B^{(0)}\rangle 
\]
\[
= \frac{s}{(2\pi)^2} \sum_{n=-\infty}^{+\infty}\int_{-\infty}^{+\infty}d\nu\ 
\int_{\delta -i\infty}^{\delta+i\infty} \frac{d\omega }{2\pi i} 
\left( \frac{s}{s_{0}}\right)^{\omega} \ \alpha_s^2 \ 
c_1(\nu,n)\, c_2(\nu,n) \ \frac{1}{\omega - \bar\alpha_s \chi(\nu,n)}
\]
\beq
= \frac{s}{(2\pi)^2} \sum_{n=-\infty}^{+\infty}\int_{-\infty}^{+\infty}d\nu\ 
\left( \frac{s}{s_{0}}\right)^{\bar\alpha_s \chi(\nu,n)} \ \alpha_s^2 \ 
c_1(\nu,n)\, c_2(\nu,n)  \; ,
\label{Ar_final_LLA}
\eeq
where
\beq
c_1(\nu,n)\equiv\langle \Phi_A^{(0)} | \ \frac{1}{\hat{\vec q}^{\:2}} \ | \nu^{(0)},n\rangle = \int d^2q \,\, \Phi_A^{(0)}(\vec q)
\frac{\left(\vec q^{\, 2}\right)^{i\nu-\frac{3}{2}}}{\pi \sqrt{2}} e^{i n \theta}\; ,
\label{IF1_proj_Born}
\eeq
\beq
c_2(\nu,n)\equiv\langle \nu^{(0)},n | \ \frac{1}{\hat{\vec q}^{\:2}} \ | \Phi_B^{(0)}\rangle = \int d^2q \,\, \Phi_B^{(0)}(-\vec q)
\frac{\left(\vec q^{\, 2}\right)^{-i\nu-\frac{3}{2}}}{\pi \sqrt{2}} e^{-i n \theta}\;.
\label{IF2_proj_Born}
\eeq
By applying the saddle point method to the $\nu$-integration, one can see that, in the large-$s$ limit, 
\beq 
\frac{\Im m_{s} {\cal A}}{s} \ \sim \ \frac{1}{\sqrt{\ln(s/s_0)}}
\left(\frac{s}{s_0}\right)^{\bar\alpha_s \, \max_{n,\nu} \chi(\nu,n)}
= \frac{1}{\sqrt{\ln(s/s_0)}}\left(\frac{s}{s_0}\right)^{\bar\alpha_s \, 4\ln 2}\;.
\eeq
Here, $\omega_P^{(0)}\equiv \bar \alpha_s\, 4 \ln 2$ is the famous BFKL Pomeron intercept in the LLA, which, strictly speaking, is not well defined because neither the scale at which $\alpha_s$ should be taken, nor the value of $s_0$ are determined.
From now on, we restrict to processes for which both impact factors possess azimuthal symmetry, {\it i.e.} $\Phi_A(\vec q)=\Phi_A(\vec q^{\:2})$, and similarly for $\Phi_B$.
In this case, the integration over the azimuthal angle in Eqs.~(\ref{IF1_proj_Born}) and~(\ref{IF2_proj_Born}) is non-vanishing only for $n=0$, thus restricting the summation over $n$ in the expression~(\ref{Ar_final_LLA}) for the amplitude to the only term with $n=0$, leading to
\beq
\Im m_{s} {\cal A}\biggl.\biggr|_{\rm LLA}
= \frac{s}{(2\pi)^2} \int_{-\infty}^{+\infty}d\nu\ 
\left( \frac{s}{s_{0}}\right)^{\bar\alpha_s \chi(\nu)} \ \alpha_s^2 \ 
c_1(\nu)\, c_2(\nu)\;,
\eeq
with
\[
\chi(\nu)\equiv \chi(\nu,0), \;\;\; c_{1,2}(\nu)\equiv c_{1,2}(\nu,0)\;.
\]
The $n=0$ sector is the only active one in the case of forward elastic or exclusive processes, such as $\gamma^*\gamma^* \to \gamma^*\gamma^*$ or $\gamma^*\gamma^* \to \rho\rho$. For forward-backward semihard processes, such as Mueller-Navelet jet production, the final-state identified objects have a definite transverse momentum, thus breaking the azimuthal symmetry and opening the way to sectors with $n\neq 0$. Nonetheless, the $n=0$ sector is always the one giving the leading contribution in the high-energy limit. 

When moving to the NLLA, impact factors and kernel must be taken at the NLO. The eigenstates $\{|\nu^{(0)}\rangle,\nu\in \mathbb{R}\}$ cease to be eigenstates
of the NLO BFKL kernel. Indeed, we have~\cite{hep-ph/9802290,Camici:1997ta} (see also
Ref.~\cite{Ivanov:2005gn})
\[
\hat K|\nu^{(0)}\rangle = \bar \alpha_s(\mu) \chi(\nu)|\nu^{(0)}\rangle
\]
\beq
+\bar \alpha_s^2(\mu_R)
\left(\chi^{(1)}(\nu)
+\frac{\beta_0}{4N_c}\chi(\nu)\ln \mu_R^2\right)|\nu^{(0)}\rangle
+ \bar
\alpha_s^2(\mu_R)\frac{\beta_0}{4N_c}\chi(\nu)\left(i\frac{\partial}{\partial \nu}
\right)|\nu^{(0)}\rangle \;,
\label{Konnu}
\eeq
where the first term in the r.h.s. represents the action of LO kernel, while the second
and the third ones stand for the diagonal and non-diagonal parts of the action of the NLO kernel. Here, $\mu_R$ is the renormalization scale, $\beta_0$ is the leading coefficient of the QCD $\beta$-function,
\[
\beta_0 = \frac{11 N_c}{3} - \frac{2 n_f}{3}\;,
\]
where $n_f$ is the number of active quark flavors; then,
\begin{equation}
\chi^{(1)}(\nu)=-\frac{\beta_0}{8\, N_c}\left(
\chi^2(\nu)-\frac{10}{3}\chi(\nu)-i\chi^\prime(\nu)
\right) + {\bar \chi}(\nu)\;,
\end{equation}
with
\[
\bar \chi(\nu)= -\frac{1}{4}\left[\frac{\pi^2-4}{3}\chi(\nu)-6\zeta(3)-
\chi^{\prime\prime}(\nu)-\frac{\pi^3}{\cosh(\pi\nu)}
\right.
\]
\beq
+ \left.
\frac{\pi^2\sinh(\pi\nu)}{2\,\nu\, \cosh^2(\pi\nu)}
\left(3+\left(1+\frac{n_f}{N_c^3}\right)\frac{11+12\nu^2}{16(1+\nu^2)}
\right) +4\phi(\nu)\right] \; ,
\eeq
\begin{equation}
\phi(\nu)\,=\,2\int\limits_0^1dx\,\frac{\cos(\nu\ln(x))}{(1+x)\sqrt{x}}
\left[\frac{\pi^2}{6}-\mbox{Li}_2(x)\right]\, , \;\;\;\;\;
\mbox{Li}_2(x)=-\int\limits_0^xdt\,\frac{\ln(1-t)}{t} \;,
\label{phi1}
\end{equation}
and $\chi^\prime(\nu)=d\chi (\nu)/d\nu$ and $\chi^{\prime\prime}
(\nu)=d^2 \chi (\nu)/d\nu^2$.

To write the amplitude with NLLA, we can proceed as in Ref.~\cite{Ivanov:2005gn}
and keep using the completeness of the LO eigenfunctions $\{|\nu^{(0)}\rangle\}$ in the expression~(\ref{Ar_op}) for the amplitude, taking the following NLO solution of the BFKL equation:
\beq
\label{GreenOperator}
\hat G_\omega=(\omega-\bar \alpha_s\hat K^{(0)})^{-1}+
(\omega-\bar \alpha_s\hat K^{(0)})^{-1}\left(\bar \alpha_s^2 \hat K^{(1)}\right)
(\omega-\bar \alpha_s \hat
K^{(0)})^{-1}+ {\cal O}\left[\left(\bar \alpha_s^2 \hat K^{(1)}\right)^2\right]
\;.
\eeq
Then, after some simple algebraic steps, one can derive the following expression
for the amplitude in NLLA:
\[
\Im m_{s} {\cal A} =\frac{s}{(2\pi)^2}
\int\limits^{+\infty}_{-\infty}
d\nu \left(\frac{s}{s_0}\right)^{\bar \alpha_s(\mu_R) \chi(\nu)} \ \alpha_s^2 \ 
c_1(\nu)c_2(\nu)
\]
\beq
\times \Biggl[1 +\bar \alpha_s(\mu_R)
\left(\frac{c^{(1)}_1(\nu)}{c_1(\nu)}
+\frac{c^{(1)}_2(\nu)}{c_2(\nu)}\right)
\label{A_basisLO}
\eeq
\[
+\bar \alpha_s^2(\mu_R)\ln\left(\frac{s}{s_0}\right)
\left(\bar
\chi(\nu)+\frac{\beta_0}{8N_c}\chi(\nu)\left[-\chi(\nu)+\frac{10}{3}
+i\frac{d\ln(\frac{c_1(\nu)}{c_2(\nu)})}{d\nu}+2\ln\mu_R^2\right]
\right)\Biggr]\;,
\]
where
\beq
c_1^{(1)}(\nu)\equiv\langle \Phi_A^{(1)} | \ \frac{1}{\hat{\vec q}^{\:2}} \ | \nu^{(0)}\rangle = \int d^2q \,\, \Phi_A^{(1)}(\vec q^{\:2})
\frac{\left(\vec q^{\, 2}\right)^{i\nu-\frac{3}{2}}}{\pi \sqrt{2}}\; ,
\label{IF1_proj_NLO}
\eeq
\beq
c_2^{(1)}(\nu)\equiv\langle \nu^{(0)} | \ \frac{1}{\hat{\vec q}^{\:2}} \ | \Phi_B^{(1)}\rangle = \int d^2q \,\, \Phi_B^{(1)}(\vec q^{\:2})
\frac{\left(\vec q^{\, 2}\right)^{-i\nu-\frac{3}{2}}}{\pi \sqrt{2}}\;.
\label{IF2_proj_NLO}
\eeq
This expression for the NLLA amplitude, and other related representations, equivalent to it within the NLLA~\cite{Ivanov:2006gt,Caporale:2015uva}, have been at the basis of numerous phenomenological studies, as already discussed in the Introduction.

\section{The Chirilli-Kovchegov eigenfunctions of the NLO BFKL kernel}
\label{sec:CK}
\begin{figure}
    \includegraphics[width=0.80\textwidth]{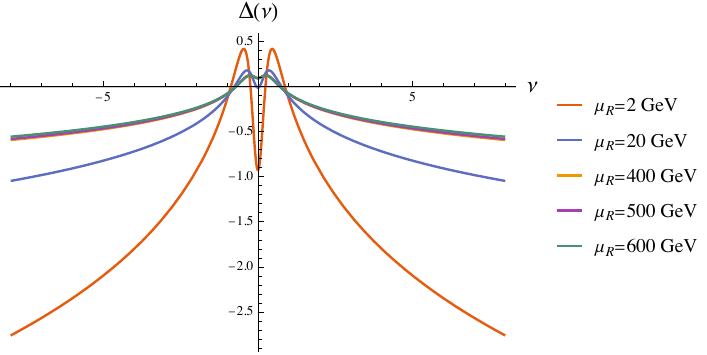}
    \caption{Behavior of the NLO eigenvalue $\Delta(\nu)$ {\it versus} $\nu$ at different values of the renormalization scale $\mu_R$.}
    \label{Deltanu}
\end{figure}

\begin{figure}
    \includegraphics[width=0.80\linewidth]{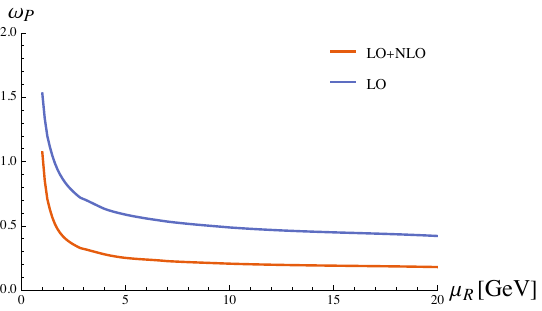}
    \caption{Behavior of the Pomeron intercept $\omega_P$, at the LO and within NLO, with the renormalization scale $\mu_R$.}
    \label{NLO_Pomeron}
\end{figure}

In Ref.~\cite{1305.1924} Chirilli and Kovchegov constructed a perturbative solution of the eigenvalue problem for the NLO BFKL kernel in the $n=0$ sector. Their findings can be summarized as follows:
\[
\hat K \, |\nu \rangle = \Delta(\nu) \, |\nu\rangle \;, \;\;\;
\Delta(\nu)= \bar \alpha_s(\mu_R) \chi(\nu) + \bar\alpha_s^2(\mu_R)
\chi_1(\nu) \;, \;\;\;
\chi_1(\nu) = \chi^{(1)}(\nu) - i\frac{\beta_0}{8\, N_c}\chi^\prime(\nu)\;,
\]
\[
\langle\vec q\, |\nu\rangle \equiv \phi_\nu(\vec q^{\:2})\;, \;\;\;
\phi_\nu (\vec q^{\:2}) = \phi_\nu^{(0)}(\vec q^{\:2}) \, \Bigg[ 1 +
    \bar{\alpha}_s(\mu_R) \, h_\nu(\vec q^{\:2}) \Biggr]\;,
\]
\beq
h_\nu(\vec q^{\:2}) \equiv A_\nu \, \ln^2\frac{\vec q^{\:2}}{\mu_R^2}  + B_\nu \, 
\ln\frac{\vec q^{\:2}}{\mu_R^2} \; ,
\label{eigenf_nu_final}
\eeq
\[
A_\nu \equiv i\, \frac{\beta_0}{8N_c} \, \frac{\chi(\nu)}{\chi'(\nu)} \;,\;\;\;\;\;
B_\nu \equiv \frac{\beta_0}{8N_c} \,  \left( 1
        - \frac{\chi(\nu) \, \chi'' (\nu)}{\chi'(\nu)^2}\right)=\frac{\beta_0}{8N_c} \,  \frac{d}{d\nu} \left(\frac{\chi(\nu)}{\chi'(\nu)}\right) \;.
\]
The NLO eigenvalue is a smooth function of $\nu$, with two degenerate maxima, symmetric with respect to $\nu=0$ (see
Fig.~\ref{Deltanu}). The NLO Pomeron intercept, which is the value taken by 
$[\Delta(\nu)]$ in these maxima, gives a negative correction to $\omega_P^{(0)}$, as shown in Fig.~\ref{NLO_Pomeron}. The presence of two degenerate maxima is responsible of the well-known numerical instabilities of the BFKL amplitude (see, {\it e.g.}, Ref.~\cite{Salam:1998tj}).


Concerning the NLO eigenfunctions, they form a complete orthonormal set in the transverse space, as in the LO case. However, differently from the LO eigenfunctions,
they have a divergence in $\nu=0$, due to the denominators with $\chi'(\nu)$, and, therefore, must be meant as distributions:
\[
A_\nu = i\, \frac{\beta_0}{8N_c} \, P \frac{\chi(\nu)}{\chi'(\nu)} 
= i\, \frac{\beta_0}{8N_c} \, \chi(\nu) P \frac{1}{\chi'(\nu)} 
\;,\;\;\;\;\;
B_\nu =\frac{\beta_0}{8N_c} \,  \frac{d}{d\nu} \left(P \frac{\chi(\nu)}{\chi'(\nu)}\right) \; ,
\]
where the symbol $P$ (which will be omitted in the rest of the paper) stands for the Cauchy principal value. 

The Chirilli-Kovchegov construction was extended in Ref.~\cite{Chirilli:2014dcb} to include the dependence on the azimuthal angle, {\it i.e.} to the sectors with $n\neq 0$~\footnote{This extension was also obtained in Ref.~\cite{DelDuca:2017peo}, where it was shown that the BFKL ladder can be evaluated order by order in the coupling in terms of generalized single-valued multiple polylogarithms.}. The result for $n \neq 0$ is~\footnote{See the Appendix for an alternative derivation.}
\beq
\hat K \, |\nu,n \rangle = \Delta(\nu,n) \, |\nu,n\rangle \;, \;\;\;
\Delta(\nu,n)= \bar \alpha_s(\mu_R) \chi(\nu,n) + \bar\alpha_s^2(\mu_R)
\chi_1(\nu,n) \;, 
\label{eigen_gen_n}
\eeq
\beq
\langle\vec q\, |\nu,n\rangle \equiv \phi_{\nu,n}(\vec q)\;, \;\;\;
\phi_{\nu,n} (\vec q) = \phi_{\nu,n}^{(0)}(\vec q) \, \Bigg[ 1 +
    \bar{\alpha}_s(\mu_R) \, h_{\nu,n}(\vec q^{\:2}) \Biggr]\;,
\eeq
\[
\phi_{\nu,n}^{(0)} \equiv \phi_\nu^{(0)}(\vec q^{\:2}) e^{i n \theta}\;, \;\;\;
\cos\theta \equiv \frac{q_x}{|\vec q \; |}\;,
\]
\beq
h_{\nu,n}(\vec q^{\:2}) \equiv A_{\nu,n} \, \ln^2\frac{\vec q^{\:2}}{\mu_R^2}  + B_{\nu,n} \,  \ln\frac{\vec q^{\:2}}{\mu_R^2} \; ,
\label{eigenf_nu_n_final}
\eeq
\[
A_{\nu,n} \equiv i\, \frac{\beta_0}{8N_c} \, \frac{\chi(\nu,n)}{\chi'(\nu,n)} \;,\;\;\;\;\;
B_{\nu,n} \equiv \frac{\beta_0}{8N_c} \,  \left( 1
        - \frac{\chi(\nu,n) \, \chi''(\nu,n)}{\chi'(\nu,n)^2}\right)
        = \frac{\beta_0}{8N_c} \,  \frac{d}{d\nu}\left( \frac{\chi(\nu,n)}{\chi'(\nu,n)}\right)\;,
\]
with 
\begin{equation}
\chi_1(\nu,n)=-\frac{\beta_0}{8\, N_c}\left(
\chi^2(\nu,n)-\frac{10}{3}\chi(\nu,n)\right) + {\bar \chi}(\nu,n)\;,
\end{equation}
\beq
\bar \chi(\nu,n)= -\frac{1}{4}\left[\frac{\pi^2-4}{3}\chi(\nu,n)-6\zeta(3)
-\chi^{\prime\prime}(\nu) +2\phi(\nu,n)+2 \phi(-\nu,n) \right.
\eeq
\[
+ \left.
\frac{\pi^2\sinh(\pi\nu)}{2\,\nu\, \cosh^2(\pi\nu)}
\left(\left(3+\left(1+\frac{n_f}{N_c^3}\right)\frac{11+12\nu^2}{16(1+\nu^2)}
\right)\delta_{n0} -\left(1+\frac{n_f}{N_c^3}\right)
\frac{1+4\nu^2}{32(1+\nu^2)}\delta_{n2}\right)\right] \; ,
\]
\beq
\phi(\nu,n)\,=\,-\int_0^1 dx \frac{x^{-1+i\nu+n/2}}{1+x} \left[
\frac{1}{2}\left(\psi'\left(\frac{n+1}{2}\right)-\zeta(2)\right)+{\rm Li}_2(x)+{\rm Li}_2(-x)
\right.
\eeq
\[
\left.
+\ln x\left(\psi(n+1)-\psi(1)+\ln(1+x)+\sum_{k=1}^\infty\frac{(-x)^k}{k+n}\right) + \sum_{k=1}^\infty\frac{x^k}{(k+n)^2}(1-(-1)^k)\right]
\]
\[
=\sum_{k=0}^\infty\frac{(-1)^{k+1}}{k+(n+1)/2+i\nu}\Biggl[\psi'(k+n+1)-\psi'(k+1)+(-1)^{k+1}(\beta'(k+n+1)+\beta'(k+1))\Biggr.
\]
\[
\Biggl.
-\frac{1}{k+(n+1)/2+i\nu}(\psi(k+n+1)-\psi(k+1))\Biggr]\;,
\]
\[
\beta'(z)=\frac{1}{4} \left[\psi'\left(\frac{z+1}{2}\right)-\psi'\left(\frac{z}{2}\right)\right]\;.
\]
In the previous expressions for $A_{\nu,n}$ and $B_{\nu,n}$, the principal-value prescription is implicit.

\section{BFKL amplitude via eigenfunctions of the NLO BFKL kernel}
\label{sec:NLO_eigen}

Writing the NLLA BFKL amplitude in terms of the NLO eigenfunctions of the kernel is as simple as in the case of the LLA amplitude written in terms of the LO eigenfunctions. Starting from Eq.~(\ref{Ar_op}) and using the completeness relation with respect to the basis $\{| \nu\rangle\}$, one can easily get, within NLLA accuracy,
\[
\Im m_{s} {\cal A} =\frac{s}{(2\pi)^2}
\int\limits^{+\infty}_{-\infty}
d\nu \left(\frac{s}{s_0}\right)^{\bar \alpha_s \chi(\nu)+{\bar \alpha_s}^2\chi_1(\nu)}
\ \alpha_s^2 \ {\tilde c_1}(\nu){\tilde c_2}(\nu)
\]
\begin{equation}
\times \Biggl[1 +\bar \alpha_s
\left(\frac{c^{(1)}_1(\nu)}{{\tilde c_1}(\nu)}
+\frac{c^{(1)}_2(\nu)}{{\tilde c_2}(\nu)}\right)\Biggr]\;,
\label{AmplNLLA}
\end{equation}
where $c_{1,2}^{(1)}$ are defined as in~(\ref{IF1_proj_NLO}), (\ref{IF2_proj_NLO}),
while
\bea
\tilde c_1(\nu)\equiv\langle \Phi_A^{(0)} | \ \frac{1}{\hat{\vec q}^{\:2}} \ | \nu\rangle &=& \int d^2q \,\, \Phi_A^{(0)}(\vec q^{\:2}) \frac{\phi_\nu(\vec q^{\:2})}{\vec q^{\:2}} \nonumber \\
&=&c_1(\nu)+\bar \alpha_s \int d^2q \,\, \Phi_A^{(0)}(\vec q^{\:2}) 
\frac{\phi^{(0)}_\nu(\vec q^{\:2})}{\vec q^{\:2}} h_\nu(\vec q^{\:2}) \; ,
\label{IF1_proj_Born_new}
\eea
\bea
\tilde c_2(\nu)\equiv\langle \nu | \ \frac{1}{\hat{\vec q}^{\:2}} \ | \Phi_B^{(0)}\rangle &=& \int d^2q \,\, \Phi_B^{(0)}(\vec q^{\:2}) 
\frac{\phi^*_\nu(\vec q^{\:2})}{\vec q^{\:2}} \nonumber \\
&=&c_2(\nu)+\bar \alpha_s \int d^2q \,\, \Phi_B^{(0)}(\vec q^{\:2}) 
\left[\frac{\phi^{(0)}_\nu(\vec q^{\:2})}{\vec q^{\:2}} h_\nu(\vec q^{\:2})\right]^* \; .
\label{IF2_proj_Born_new}
\eea
Recalling the expressions for $h_\nu$, given in~(\ref{eigenf_nu_final}), and for
$\phi_\nu^{(0)}(\vec q^{\:2})$, given in~(\ref{LO_eigen}), 
we can rewrite $\tilde c_{1,2}^{(0)}(\nu)$ as
\beq
\tilde c_1(\nu) = \left\{1 + \bar \alpha_s 
\left[ A_\nu \left(- i \frac{d}{d\nu} - \ln \mu_R^2\right)^2 +
B_\nu \left(- i \frac{d}{d\nu} - \ln \mu_R^2\right)\right] \right\}c_1(\nu) \;,
\label{c1_new}
\eeq
\beq
\tilde c_2(\nu) = \left\{1 + \bar \alpha_s 
\left[ - A_\nu \left( + i \frac{d}{d\nu} - \ln \mu_R^2\right)^2 +
B_\nu \left( + i \frac{d}{d\nu} - \ln \mu_R^2\right)\right] \right\}c_2(\nu) \;,
\label{c2_new}
\eeq
so that, within NLLA accuracy, the amplitude takes the form
\[
\Im m_{s} {\cal A} =\frac{s}{(2\pi)^2}
\int\limits^{+\infty}_{-\infty}
d\nu \left(\frac{s}{s_0}\right)^{\bar \alpha_s \chi(\nu)+{\bar \alpha_s}^2\chi_1(\nu)}
\ \alpha_s^2 \ c_1(\nu) c_2(\nu)
\]
\beq
\times \Biggl[1 +\bar \alpha_s
\left(\frac{c^{(1)}_1(\nu)}{c_1(\nu)}
+\frac{c^{(1)}_2(\nu)}{c_2(\nu)}\right) \Biggr.
\label{AmplNLLA_final}
\eeq
\[
+\bar \alpha_s \Biggl(-A_\nu \left(\frac{c_1''(\nu)}{c_1(\nu)}
-\frac{c_2''(\nu)}{c_2(\nu)}\right)
-i B_\nu \, \frac{d \ln\left(\frac{c_1(\nu)}{c_2(\nu)}\right)}{d\nu} + 2 \ln\mu_R^2 \left( i A_\nu \frac{d\ln c_1(\nu) c_2(\nu)}{d\nu}-B_\nu\right)\Biggr) \Biggr]\;.
\]
We observe that the integrand of this expression contains terms singular in $\nu=0$, due to the presence of $\chi'(\nu)$ in the denominators of $A_\nu$ and $B_\nu$ (see their definition in~(\ref{eigenf_nu_final})). A simple inspection of~(\ref{AmplNLLA_final}) and~(\ref{A_basisLO}) shows that they are equivalent in the NLLA if the following two expressions,
\[
I = \int_{-\infty}^{+\infty} d\nu \, \left(\frac{s}{s_0}\right)^{\bar \alpha_s \chi}
c_1 c_2 \left[i \frac{\chi}{\chi'}\left(\frac{c_2''}{c_2}-\frac{c_1''}{c_1}\right)
-i \frac{d}{d\nu}\left(\frac{\chi}{\chi'}\right) \frac{d}{d\nu}\ln\frac{c_1}{c_2}
\right.
\]
\[
\left. -2 \ln \mu_R^2 \left( \frac{\chi}{\chi'} \frac{d}{d\nu}\ln(c_1 c_2)
+ \frac{d}{d\nu}\left(\frac{\chi}{\chi'}\right) \right)\right]
\]
and
\[
J = \int_{-\infty}^{+\infty} d\nu \, \left(\frac{s}{s_0}\right)^{\bar \alpha_s \chi}
\, c_1 c_2 \, \bar\alpha_s \, \chi\, \ln\left(\frac{s}{s_0}\right) 
\left(  i\frac{d}{d\nu}\ln\frac{c_1}{c_2} + 2 \ln\mu_R^2 \right) \;,
\]
coincide. Using the identities 
\[
c_1 c_2\, \left[\frac{\chi}{\chi'}\left(\frac{c_2''}{c_2}-\frac{c_1''}{c_1}\right)
- \frac{d}{d\nu}\left(\frac{\chi}{\chi'}\right) \frac{d}{d\nu}\ln\frac{c_1}{c_2}\right]
= -\,\frac{d}{d\nu}\left( c_1 c_2 \, \frac{\chi}{\chi'}\, \frac{d}{d\nu} \ln\frac{c_1}{c_2}\right)\;,
\]
\[
c_1 c_2 \, \left[\frac{\chi}{\chi'} \frac{d}{d\nu}\ln(c_1 c_2)
+ \frac{d}{d\nu}\left(\frac{\chi}{\chi'}\right)\right] = \frac{d}{d\nu}\left( c_1 c_2
\frac{\chi}{\chi'}\right)\;,
\]
we can rewrite $I$ as
\[
I = \int_{-\infty}^{+\infty} d\nu \, \left(\frac{s}{s_0}\right)^{\bar \alpha_s \chi}
\, \frac{d}{d\nu} \left[ - c_1 c_2 \, \frac{\chi}{\chi'}\, \left( i \frac{d}{d\nu} \ln\frac{c_1}{c_2} +2 \ln\mu_R^2\right)\right]\;.
\]
Then, the equivalence with $J$ results from 
\[
\int_{-\infty}^{+\infty} d\nu \, \left(\frac{s}{s_0}\right)^{\bar \alpha_s \chi}
\, \bar\alpha_s \ln\left(\frac{s}{s_0}\right) \, f(\nu)
= \int_{-\infty}^{+\infty} d\nu \, \left(\frac{s}{s_0}\right)^{\bar \alpha_s \chi}
\, \bar\alpha_s \ln\left(\frac{s}{s_0}\right) \, f(\nu)\, \chi' P\frac{1}{\chi'}
\]
\[
=-\int_{-\infty}^{+\infty} d\nu \, \left(\frac{s}{s_0}\right)^{\bar \alpha_s \chi}\,
\frac{d}{d\nu} \left(P\frac{f(\nu)}{\chi'(\nu)}\right)\;,
\]
where an integration by parts was performed in the last step. 

The NLLA equivalence between~(\ref{AmplNLLA_final}) and~(\ref{A_basisLO}) proves that the singularities at $\nu=0$ in the integrand of~(\ref{AmplNLLA_final}) are harmless, since they can be removed by an integration by parts bringing the amplitude back to the form~(\ref{A_basisLO}), which is manifestly exempt from singular terms. However, in practical applications one might prefer to work with the expression~(\ref{AmplNLLA_final}), because the different treatment of terms beyond the NLLA~\footnote{We stress that terms beyond the NLLA are unavoidably present in both representations; as an example, terms of the form $\bar \alpha_s^2 \ln (s/s_0)$ contain the piece $\bar \alpha_s^2 \ln s_0$ which is evidently next-to-NLLA.} could improve the numerical stability of the result. In this case, the singular terms in the integrand (contained in $A_\nu$ and $B_\nu$) should be taken with the principal-value prescription. In numerical implementations, it can be simply realized by the replacement
\beq
P\frac{1}{\chi'(\nu)} \to {\cal R}e \, \Bigg[\frac{1}{\chi'(\nu)+i \epsilon}\Bigg]\;,
\label{VP}
\eeq
with $\epsilon$ an infinitesimal parameter. 

The extension of the results of the present Section to inclusive forward-backward processes, where also the kernel eigenfunctions with $n\neq 0$ can play a role, is straightforward.

\section{Application to the NLLA amplitude for the electroproduction of two light vector mesons}
\label{sec:rho_prod}
\begin{figure}
\includegraphics[scale=0.5]{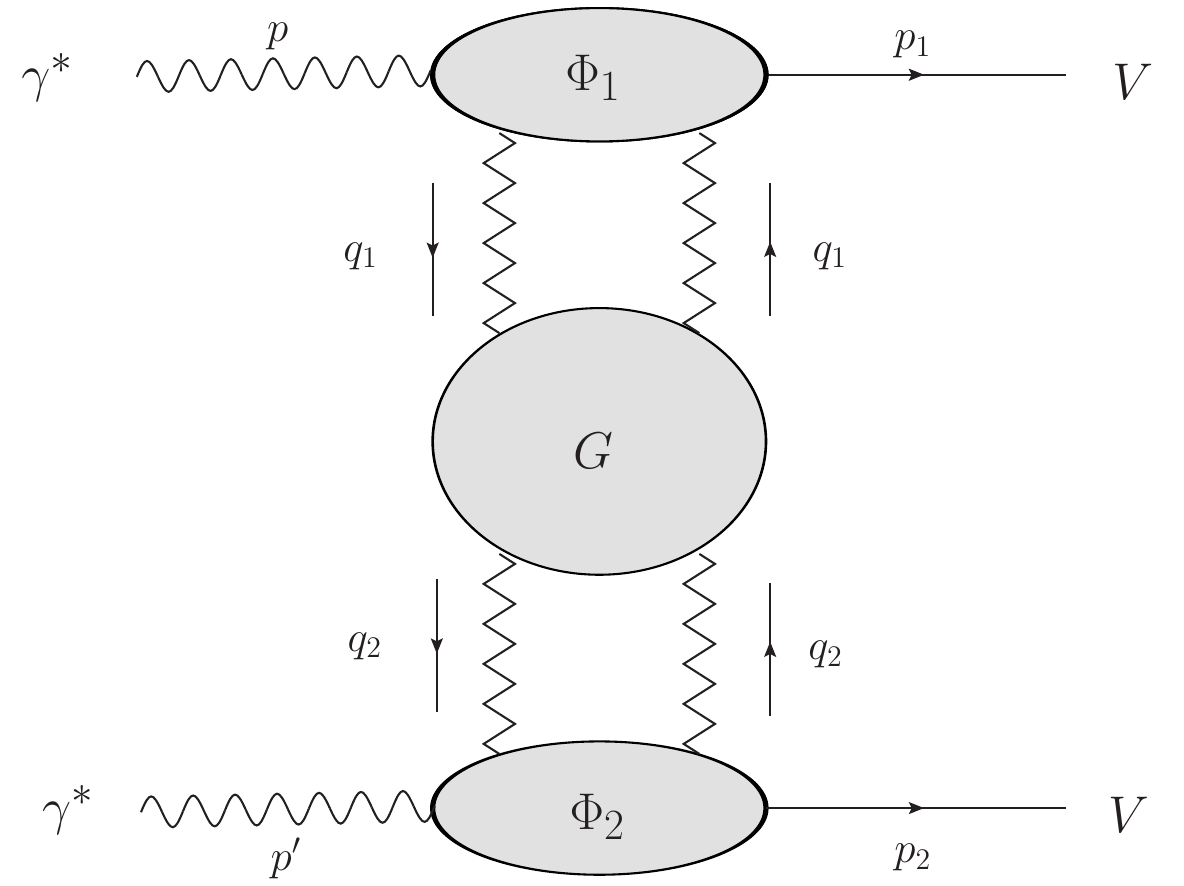}
\caption{Schematic representation of the BFKL amplitude for the forward electroproduction of two light vector mesons.}
\label{fig:rho}
\end{figure}

In this section, we consider the impact factor for the production of a light vector meson to construct an NLLA representation of a specific amplitude, using the basis of NLO kernel eigenfunctions. Instead of the more famous diffractive vector meson production, which has long been investigated in the context of small-$x$ physics~\cite{Kowalski:2006hc,Anikin:2011sa,Besse:2012ia,Besse:2013muy,Bolognino:2018rhb,Bolognino:2019pba,Bolognino:2021niq,Mantysaari:2020lhf}, we consider the amplitude for the electroproduction of two light vector mesons,  $\gamma^* + \gamma^* \to V + V$, with $V=\rho^0, \omega, \phi$. In this way, the only non-perturbative ingredient is represented by a leading-twist distribution amplitude (DA), and the ambiguities related to the parametrization of the proton impact factor are avoided. We consider only the dominant (leading twist) contribution to this process, involving the transition from a longitudinally polarized virtual photon to a longitudinally polarized vector meson, which is known at NLO~\footnote{The transversally polarized photon to longitudinally polarized meson transition amplitude, suppressed by one power of the hard scale, is known at the NLO only in the Wilson line formalism~\cite{Boussarie:2016bkq}. By contrast, all transitions involving a transversally polarized meson, also suppressed by at least one power of the hard scale, are known only at the leading order~\cite{Anikin:2009hk,Anikin:2009bf,Boussarie:2024pax,Boussarie:2024bdo}.}. We focus on the kinematic regime when the photon virtualities are much larger than $\Lambda_{\rm QCD}^2$ and, at the same time, much smaller than the squared center-of-mass-energy (see Fig.~\ref{fig:rho}). This amplitude has been the first one ever written within the full NLLA BFKL approach~\cite{Ivanov:2005gn}. One of the key observations of Ref.~\cite{Ivanov:2005gn}, confirmed by a subsequent extended numerical analysis~\cite{Ivanov:2006gt}, was that NLO corrections, coming both from the kernel and the impact factors, are large and negative with respect to the corresponding LO contributions, thus signaling a severe instability of the perturbative series. As a consequence, the amplitude exhibits a strong dependence on the energy scales $s_0$ and $\mu_R$, through the next-to-NLLA terms unavoidably entering its expression. The search of values of $s_0$ and $\mu_R$ for which the amplitude is least sensitive to their variation, in the spirit of the principle of minimum sensitivity (PMS) method~\cite{Stevenson:1981vj}, has lead to "optimal" values of $s_0$ and $\mu_R$ much above the intrinsic hard scales of the process, provided by the photon virtualities. The idea here is to check whether a representation based on the NLO kernel eigenfunctions brings along a shift of the optimal scales towards their "natural" values. \\

Let us first fix the kinematic of the process under consideration
\begin{equation}
\gamma^*(p) \: \gamma^*(p')\to V(p_1) \:V(p_2) \;.
\end{equation}
Here, neglecting the meson masses, $p_1$ and $p_2$ are taken as Sudakov vectors satisfying $p_1^2=p_2^2=0$ and $2(p_1 p_2)=s$; the virtual photon momenta are instead
\begin{equation}
p =\alpha p_1-\frac{Q_1^2}{\alpha s} p_2 \;, \hspace{2cm}
p'=\alpha^\prime p_2-\frac{Q_2^2}{\alpha^\prime s} p_1 \;,
\end{equation}
so that the photon virtualities turn to be $p^2=-Q_1^2$ and $(p')^2=-Q_2^2$.
As anticipated, we consider the semihard regime, 
\begin{equation}
s\gg Q^2_{1,2}\gg \Lambda^2_{QCD} \, ,
\end{equation}
and
\begin{equation}
\alpha=1+\frac{Q_2^2}{s}+{\cal O}(s^{-2})\, , \quad
\alpha^\prime =1+\frac{Q_1^2}{s}+{\cal O}(s^{-2})\, .
\end{equation}
Moreover, we restrict to the {\it forward} case, {\it i.e.} when the transverse momenta of produced $V$ mesons are zero or when the variable $t=(p_1-p)^2$ takes its maximal value $t_0=-Q_1^2Q_2^2/s+{\cal O}(s^{-2})$. 

The impact factors for this process were calculated in~\cite{Ivanov:2004pp} and can be presented as an expansion in $\alpha_s$,
\begin{equation}
\label{ImpFact}
\Phi_{1,2}(\vec q)= \alpha_s \,
D_{1,2}\left[C^{(0)}_{1,2}(\vec q^{\,\, 2})+\bar\alpha_s
C^{(1)}_{1,2}(\vec
q^{\,\, 2})\right] \, , \quad D_{1,2}=-\frac{4\pi e_q  f_V}{N_c Q_{1,2}}
\sqrt{N_c^2-1}\; ,
\end{equation}
where $f_V$ is the meson coupling constant ($f_{\rho}\approx
200\, \rm{ MeV}$) and $e_q$ should be replaced by $e/\sqrt{2}$, $e/(3\sqrt{2})$
and $-e/3$ for the case of $\rho^0$, $\omega$ and $\phi$ meson production,
respectively~\footnote{Please notice the slight change of notation for impact factors, with respect to~(\ref{phi_expansion}):
$\Phi_{A,B} \to \Phi_{1,2}$.}.

In the collinear factorization approach, in which the meson is produced collinearly with respect to the initial virtual photon, the meson transition impact factor is given as a convolution of the hard scattering amplitude for the production of a collinear quark-antiquark pair with the meson DA. The integration variable in this convolution is the fraction $z$
of the meson momentum carried by the quark, while $\bar z\equiv 1-z$ is
the momentum fraction carried by the antiquark:
\begin{equation}
\label{C0}
C^{(0)}_{1,2}(\vec q^{\,\, 2})=\int\limits^1_0 dz \,
\frac{\vec q^{\,\, 2}}{\vec q^{\,\, 2}+z \bar zQ_{1,2}^2}\phi_\parallel (z)
\, .
\end{equation}
The NLO correction to the hard scattering amplitude, for a photon with virtuality equal to $Q^2$, is defined as follows:
\begin{equation}
C^{(1)}(\vec q^{\,\, 2})=\frac{1}{4 N_c}\int\limits^1_0 dz \,
\frac{\vec q^{\,\, 2}}{\vec q^{\,\, 2}+z \bar zQ^2}[\tau(z)+\tau(1-z)]
\phi_\parallel (z)
\; ,
\end{equation}
with $\tau(z)$ given by Eq.~(40) of Ref.~\cite{Ivanov:2005xc}.
$C^{(1)}_{1,2}(\vec q^{\,\, 2})$ are given by the previous expression with
$Q^2$ replaced everywhere in the integrand by $Q^2_1$ and $Q^2_2$,
respectively. 
As in Ref.~\cite{Ivanov:2005gn}, we use the distribution amplitude in its asymptotic form $\phi^{\rm as}_\parallel(z)=6z(1-z)$. Integrating over $z$ in~(\ref{C0}), we obtain, for a photon with virtuality $Q^2$, the following result:
\begin{equation}
    C^{(0)}\,\left(\alpha=\frac{\vec q^{\,\, 2}}{Q^2}\right)\,
=\,6\, \alpha \left[1-\, \frac{\alpha}{c}\,
\ln\frac{2c+1}{2c-1}
\right]\, ,
\end{equation}
where $c=\sqrt{\alpha+1/4}\,$. $C^{(0)}_{1,2}$ are given by the previous expression with $Q^2$ replaced by $Q^2_1$ and $Q^2_2$, respectively.
For the NLO term $C^{(1)}_{1,2}(\vec
q^{\:2})$ the integration over $z$ can be performed by a numerical calculation.

When the basis of LO kernel eigenfunctions is used, the NLLA amplitude takes the form
(compare with~(\ref{A_basisLO}))
\begin{equation}
\frac{\Im m_s {\cal A}}{D_1D_2}\biggr|_{\rm LO \ basis}=\frac{s}{(2\pi)^2}
\int\limits^{+\infty}_{-\infty}
d\nu \left(\frac{s}{s_0}\right)^{\bar \alpha_s \chi(\nu)}
\alpha_s^2 c_1(\nu)c_2(\nu) \notag
\end{equation}
\begin{equation}
\times \Biggl[1 +\bar \alpha_s
\left(\frac{c^{(1)}_1(\nu)}{c_1(\nu)}
+\frac{c^{(1)}_2(\nu)}{c_2(\nu)}\right) \notag 
\end{equation}
\begin{equation}
+\bar \alpha_s^2\ln\left(\frac{s}{s_0}\right)
\left(\bar
\chi(\nu)+\frac{\beta_0}{8N_c}\chi(\nu)\left[-\chi(\nu)+\frac{10}{3}
+i\frac{d\ln(\frac{c_1(\nu)}{c_2(\nu)})}{d\nu}+2\ln\mu_R^2\right]
\right)\Biggr]\;,
\label{A_0}
\end{equation}
where
\begin{equation}
c_{1,2}(\nu)=\int d^2\vec q \,\, C_{1,2}^{(0)}(\vec q^{\, 2})
\frac{\left(\vec q^{\, 2}\right)^{\pm i\nu-\frac{3}{2}}}{\pi \sqrt{2}} \;,\;\;\;\;\;
c_{1,2}^{(1)}(\nu)=\int d^2\vec q \,\, C_{1,2}^{(1)}(\vec q^{\, 2})
\frac{\left(\vec q^{\, 2}\right)^{\pm i\nu-\frac{3}{2}}}{\pi \sqrt{2}}\;.
\label{Coeffi}
\end{equation}
In the case of the LO coefficients, the integration can be performed analytically, giving
\beq
c_{1,2}(\nu) = \frac{\left(Q^2_{1,2}\right)^{\pm i\nu-\frac{1}{2}}}{\sqrt{2}}
\frac{\Gamma^2 [\frac{3}{2}\pm i\nu]}{\Gamma [3\pm 2i\nu]}
\frac{6\pi}{\cosh (\pi \nu)}\; ,
\eeq
whereas we have to resort to a numerical integration to get $c_{1,2}^{(1)}(\nu)$.
Moreover, we have also that
\beq
i\frac{d\ln(\frac{c_1(\nu)}{c_2(\nu)})}{d\nu}=2\Biggl[
\psi(3+2i\nu)+\psi(3-2i\nu)-\psi\left(\frac{3}{2}+i\nu\right) 
-\psi\left(\frac{3}{2}-i\nu\right)-\ln\left(Q_1Q_2\right)
\Biggr]\;.
\eeq

When instead the basis of NLO kernel eigenfunctions is used, the NLLA amplitude takes the form (compare with~(\ref{AmplNLLA}))
\[
\frac{\Im m_s{\cal A}}{D_1D_2}\biggr|_{\rm NLO \ basis}=\frac{s}{(2\pi)^2}
\int\limits^{+\infty}_{-\infty}
d\nu \left(\frac{s}{s_0}\right)^{\bar \alpha_s \chi(\nu)+{\bar \alpha_s}^2\chi_1(\nu)}
\alpha_s^2 \tilde c_1(\nu)\tilde c_2(\nu) 
\]
\beq
\times \Biggl[1 +\bar \alpha_s
\left(\frac{c^{(1)}_1(\nu)}{\tilde c_1(\nu)}
+\frac{c^{(1)}_2(\nu)}{\tilde c_2(\nu)}\right)\Biggr]\; ,
\label{A_1}
\eeq
with
\beq
\tilde c_1(\nu)= \int d^2q \,\, C_1^{(0)}(\vec q^{\:2}) \frac{\phi_\nu(\vec q^{\:2})}{\vec q^{\:2}} = c_1(\nu)+\bar \alpha_s \int d^2q \,\, C_1^{(0)}(\vec q^{\:2}) 
\frac{\phi^{(0)}_\nu(\vec q^{\:2})}{\vec q^{\:2}} h_\nu(\vec q^{\:2}) \; ,
\label{New-c1}
\eeq
\beq
\tilde c_2(\nu)= \int d^2q \,\, C_2^{(0)}(\vec q^{\:2}) 
\frac{\phi^*_\nu(\vec q^{\:2})}{\vec q^{\:2}} 
= c_2(\nu)+\bar \alpha_s \int d^2q \,\, C_2^{(0)}(\vec q^{\:2}) 
\left[\frac{\phi^{(0)}_\nu(\vec q^{\:2})}{\vec q^{\:2}} h_\nu(\vec q^{\:2})\right]^* \; ,
\label{New-c2}
\eeq
so that the relation between $\tilde c_{1,2}(\nu)$ and $c_{1,2}(\nu)$ is exactly as in 
Eqs.~(\ref{c1_new}) and~(\ref{c2_new}). 

We have already shown in the general case that the amplitudes written in the two basis are equivalent which each other within the NLLA. Since they embed terms beyond the NLLA in a different way, they may exhibit a different sensitivity to the scales $s_0$ and $\mu_R$. To check this possibility, we want to compare numerically the two forms of the amplitude, in the simple case in which the two photon virtualities are taken the same, $Q_1^2=Q_2^2\equiv Q^2$. 

\subsection{Standard representation}
We study the behavior of the amplitude expressed in terms of the LO basis starting from the "standard" representation of (\ref{A_0}). This representation is implemented without any relevant manipulation. In particular, as a preliminary step, we isolate the terms depending on $s_0$ and $\beta_0$ in the NLO part of the impact factors~\cite{Ivanov:2005xc},
\[
C^{(1)}(\vec q^{\,\,2})=\int\limits^1_0 dz \,
\frac{\vec q^{\,\, 2}}{\vec q^{\,\, 2}+z \bar zQ^2}\phi_\parallel (z) 
\]
\begin{equation}
\times \left[
\frac{1}{4}\ln\left(\frac{s_0}{Q^2}\right)\ln\left(\frac{(\alpha+z\bar
z)^4}{\alpha^2 z^2\bar z^2}\right)+\frac{\beta_0}{4N_c}\left(
\ln\left(\frac{\mu_R^2}{Q^2}\right)+\frac{5}{3}-\ln \alpha\right)
+\dots \right] \;,
\label{separation1}
\end{equation}
for which the projection onto LO eigenfunctions can be done analytically, and write~\cite{Ivanov:2005gn}
\begin{equation}
\frac{c^{(1)}_{1}(\nu)}{c_{1}(\nu)}+\frac{c^{(1)}_{2}(\nu)}{c_{2}(\nu)}=\ln\left(\frac{s_0}{Q^2}\right)\chi(\nu)
+\frac{\beta_0}{2N_c}\left[\ln\left(\frac{\mu_R^2}{Q^2}\right)+\frac{5}{3}\right. \notag
\end{equation}
\begin{equation}
+\left.
\psi(3+2i\nu)+\psi(3-2i\nu)-\psi\left(\frac{3}{2}+i\nu\right)
-\psi\left(\frac{3}{2}-i\nu\right)
\right] 
\end{equation}
\[
+ \frac{\bar c^{(1)}_{1}(\nu)}{c_{1}(\nu)}+\frac{\bar c^{(1)}_{2}(\nu)}{c_{2}(\nu)}\;,
\]
where $\bar c_{1,2}^{(1)}$ is the part of $c_{1,2}^{(1)}$ not depending on $s_0$ and $\beta_0$, {\it i.e.} the contribution coming from the dots in~(\ref{separation1}), which is evaluated numerically. The final form of the implemented amplitude in the LO basis, called ${\cal A}_0$, becomes:
\begin{equation}
\frac{Q^2}{D_1 D_2} \frac{\Im m_s {{\cal A}_0}}{s} =
\frac{Q^2}{(2\pi)^2}  \alpha_s^2  \notag \\
\end{equation}
\begin{equation}
\label{NLLA_A0}
\times \int\limits^{+\infty}_{-\infty}
d\nu \left(\frac{s}{s_0}\right)^{\bar \alpha_s(\mu_R) \chi(\nu)} \ 
c_1(\nu)c_2(\nu)
\Biggl[1 +\bar \alpha_s(\mu_R)
\Bigg[\ln\left(\frac{s_0}{Q^2}\right)\chi(\nu)
\notag
\end{equation}
\begin{equation}
+\frac{\beta_0}{2N_c}\left(\ln\left(\frac{\mu_R^2}{Q^2}\right)+f(\nu)
\right) + \frac{\bar c^{(1)}_{1}(\nu)}{c_{1}(\nu)}+\frac{\bar c^{(1)}_{2}(\nu)}{c_{2}(\nu)}\Bigg]+\bar \alpha_s^2(\mu_R)\ln\left(\frac{s}{s_0}\right)
\notag
\end{equation}
\begin{equation}
\times \left(\bar
\chi(\nu)+\frac{\beta_0}{8N_c}\chi(\nu)\left[-\chi(\nu)+\frac{10}{3}
+i\frac{d\ln(\frac{c_1(\nu)}{c_2(\nu)})}{d\nu}+2\ln\mu_R^2\right]
\right)\Biggr]\;,
\end{equation}
where
\begin{equation}
    f(\nu)=\frac{5}{3}+\psi(3+2i\nu)+\psi(3-2i\nu)-\psi\left(\frac{3}{2}+i\nu\right)
-\psi\left(\frac{3}{2}-i\nu\right)\;.
\end{equation}

The standard representation ${\cal A}_0$ is then compared with its correspondent form expressed in terms of the NLO basis, given by (\ref{A_1}). After some trivial manipulations, this latter amplitude, called ${\cal A}_1$, is implemented in the following way:
\begin{equation}
\frac{Q^2}{D_1 D_2} \frac{\Im m_s {{\cal A}_1}}{s} =
\frac{Q^2}{(2\pi)^2}  \alpha_s^2  \notag \\
\end{equation}
\begin{equation}
\label{NLLA_A1}
\times \int\limits^{+\infty}_{-\infty}
d\nu \left(\frac{s}{s_0}\right)^{\bar \alpha_s(\mu_R) \chi(\nu)+{\bar \alpha_s}^2\chi_1(\nu)} \ 
c_1(\nu)c_2(\nu)
\Biggl[1 +\bar \alpha_s(\mu_R)
\Bigg(A_\nu\Bigg(f_1(\nu)-2\ln\Bigg(\frac{\mu_R^2}{Q^2}\Bigg)f_2(\nu)\Bigg)
\notag
\end{equation}
\begin{equation}
+B_\nu\Bigg(-2\Bigg(f(\nu)-\frac{5}{3}\Bigg)-2\ln\Bigg(\frac{\mu_R^2}{Q^2}\Bigg)\Bigg)+\frac{\bar c^{(1)}_{1}(\nu)}{\tilde c_{1}(\nu)}+\frac{\bar c^{(1)}_{2}(\nu)}{\tilde c_{2}(\nu)}+f_3(\nu)\Bigg)\Bigg],
\end{equation}
where
\begin{equation}
    f_1(\nu)=4\Bigg(\psi(i\nu+\frac{3}{2})-\psi(2i\nu+3)\Bigg)^2 + \Bigg(\psi(i\nu+\frac{1}{2}) - \psi(-i\nu+\frac{1}{2})\Bigg) \Bigg(4\psi(i\nu+\frac{3}{2}) \notag \\
\end{equation}
\begin{equation}
 +4 \psi (-i\nu+\frac{3}{2}) - 4\psi(2i\nu+3) -4\psi(-2i\nu+3)\Bigg) - 4 \Bigg(\psi(-i\nu+\frac{3}{2}) - \psi (-2i\nu+3)\Bigg)^2 \notag\\
 \end{equation}
 \begin{equation}
 + 2\psi'(i\nu+\frac{3}{2}) - 2\psi' (-i\nu+\frac{3}{2}) - 4\psi' (2i\nu+3) + 4\psi' (-2i\nu+3)\; , \notag \\
\end{equation}
\begin{equation}
    f_2(\nu)= 2\psi(i\nu+\frac{3}{2}) -2\psi (2i\nu+3)- 2\psi(-i\nu+\frac{3}{2})  + 2\psi (-2i\nu+3) \notag \\
\end{equation}
\begin{equation}
    + 2\psi(i\nu+\frac{1}{2}) - 2\psi(-i\nu+\frac{1}{2}),\notag
\end{equation}
\begin{equation}
    f_3(\nu)=\Bigg(\frac{1}{2}\ln\left(\frac{s_0}{Q^2}\right)\chi(\nu)
+\frac{\beta_0}{4N_c}\left(\ln\left(\frac{\mu_R^2}{Q^2}\right)+f(\nu)\right)\Bigg) \notag \\
\end{equation}
\begin{equation}
\times \Bigg(\frac{1}{1+\bar\alpha_s(A_\nu I_1+B_\nu I_2)}+\frac{1}{1+\bar\alpha_s(A^*_\nu I_1^*+B^*_\nu I^*_2)}\Bigg).
 \end{equation}

We notice that, in this expression, some terms beyond the next-to-NLA are
included through the use of ${\tilde c_{1,2}}$ at the denominator.
As before, the terms that are independent from $s_0$ and $\beta_0$ are evaluated numerically. The remaining contributions are given by $f_3(\nu)$, in which $I_1$ and $I_2$ correspond to the additional terms in the definition of the projections of the impact factors on the NLO basis with respect to the ones on the LO basis. In fact, substituting (\ref{eigenf_nu_final}) in (\ref{New-c1}) and (\ref{New-c2}), we can write
 \begin{equation}
\tilde c_1(\nu)= c_1(\nu)\Bigg(1+\bar \alpha_s(A_\nu I_1+B_\nu I_2)\Bigg) \notag,
 \end{equation}
\begin{equation}
\tilde c_2(\nu)  
= c_2(\nu)\Bigg(1+\bar \alpha_s (A^*_\nu I^*_1+B^*_\nu I^*_2)\Bigg).
\end{equation}
The integrals
\begin{equation}
 \int d^2q \,\, C_1^{(0)}(\vec q^{\:2}) 
\frac{\phi^{(0)}_\nu(\vec q^{\:2})}{\vec q^{\:2}} \ln^2\frac{\vec q^{\:2}}{\mu_R^2}, \ \ \ \ \ \ \ \    
 \int d^2q \,\, C_1^{(0)}(\vec q^{\:2}) 
\frac{\phi^{(0)}_\nu(\vec q^{\:2})}{\vec q^{\:2}}\ln\frac{\vec q^{\:2}}{\mu_R^2}\; ,
\end{equation}
can be solved analytically and their results are proportional to $c_1(\nu)$ (or $c_2(\nu)$ considering the complex conjugates). Factoring out $c_1(\nu)$ ($c_2(\nu)$) from these results, what remains corresponds respectively to $I_1$ for the first integral and $I_2$ for the second one ($I^*_1$ and $I^*_2$ for the complex conjugates). Finally, the implementation of (\ref{NLLA_A1}) requires to take into account the principal-value prescription in the definition of the NLO eigenfunctions, addressed in Section~\ref{sec:CK}, which we do through the replacement in~(\ref{VP}).

\begin{figure*}
            \includegraphics[width=.5\textwidth]{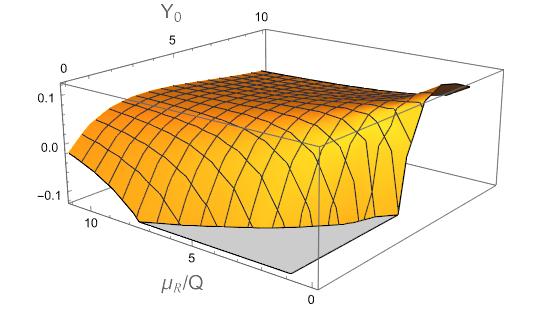}\hfill
            \includegraphics[width=.5\textwidth]{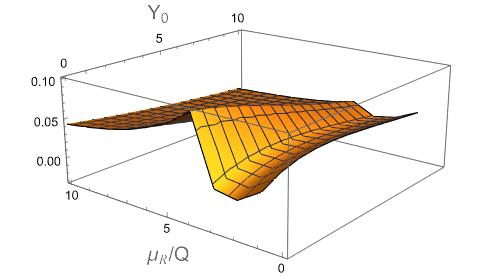}\hfill
            \caption{For $Y=10$, ${\cal I}m_s ({{\cal A}_0})Q^2/(s \, D_1 D_2)$ on the left and ${\cal I}m_s ({{\cal A}_1})Q^2/(s \, D_1 D_2)$ on the right.}
            \label{Y=10}
        \end{figure*}
\begin{figure*}
            \includegraphics[width=.5\textwidth]{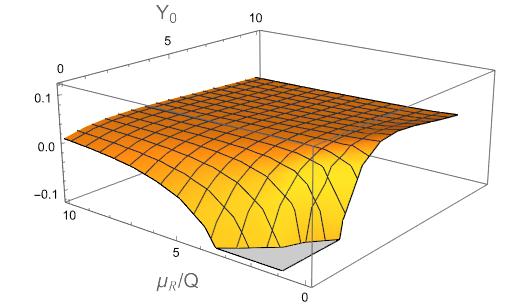}\hfill
            \includegraphics[width=.5\textwidth]{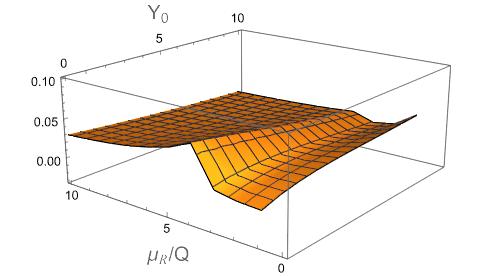}\hfill
            \caption{For $Y=6$, ${\cal I}m_s ({{\cal A}_0})Q^2/(s \, D_1 D_2)$ on the left and ${\cal I}m_s ({{\cal A}_1})Q^2/(s \, D_1 D_2)$ on the right.}
            \label{Y=6}
        \end{figure*}
\begin{figure*}
            \includegraphics[width=.5\textwidth]{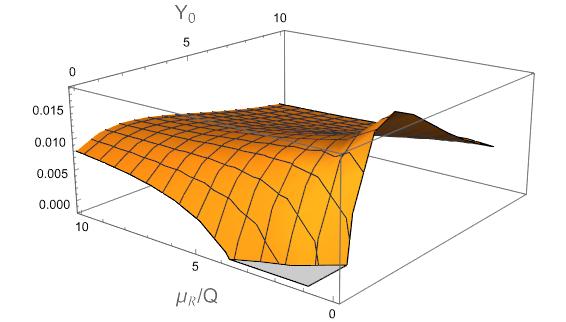}\hfill
            \includegraphics[width=.5\textwidth]{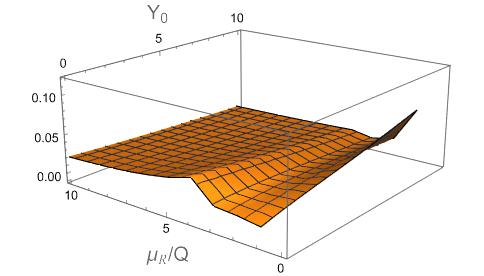}\hfill
            \caption{For $Y=2$, ${\cal I}m_s ({{\cal A}_0})Q^2/(s \, D_1 D_2)$ on the left and ${\cal I}m_s ({{\cal A}_1})Q^2/(s \, D_1 D_2)$ on the right.}
            \label{Y=2}
        \end{figure*}
\begin{figure}
            \includegraphics[width=1\textwidth]{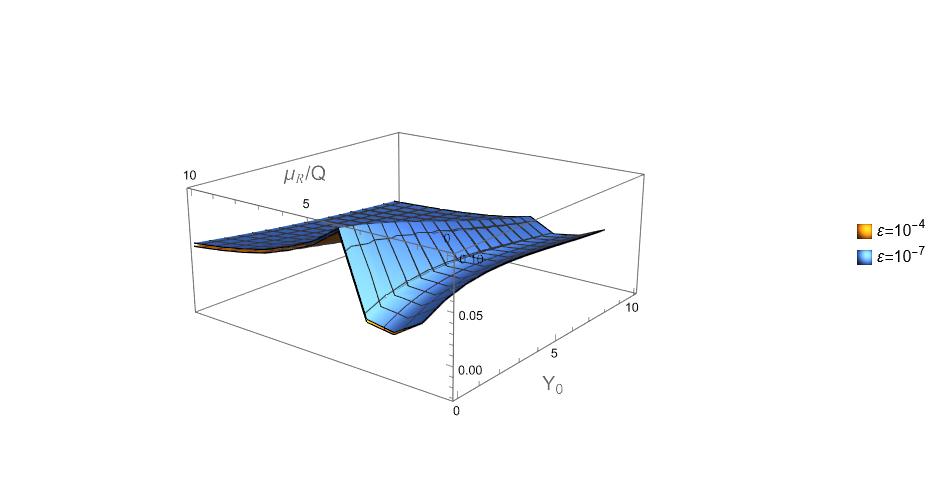}\hfill
            \includegraphics[width=1\textwidth]{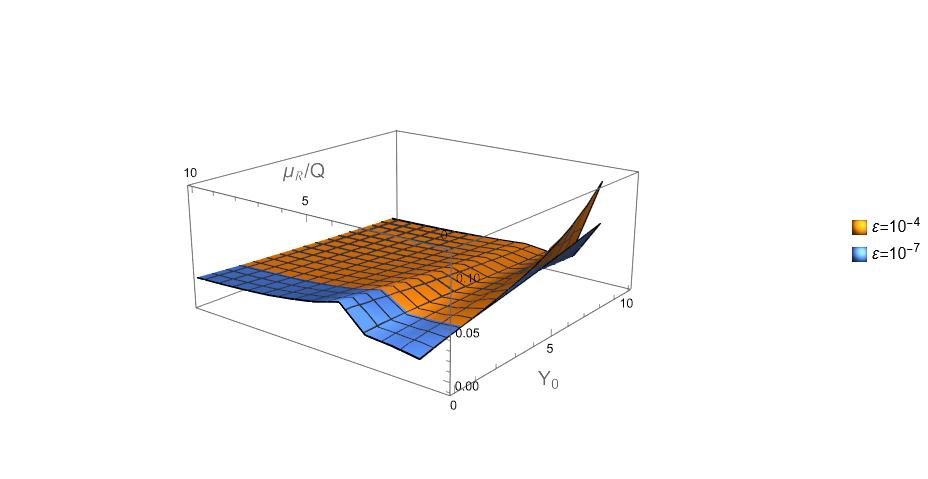}\hfill
            \caption{${\cal I}m_s ({{\cal A}_1})Q^2/(s \, D_1 D_2)$ with $\epsilon=10^{-4},10^{-7}$; for $Y=10$ (up) and $Y=2$ (down), at $Q^2= 24$ GeV$^2$.}
            \label{epsilon}
\end{figure}

In Figs~\ref{Y=10}, \ref{Y=6}, \ref{Y=2} we present
the behavior of ${\cal I}m_s ({{\cal A}_0})Q^2/(s \, D_1 D_2)$ (left plots) and ${\cal I}m_s ({{\cal A}_1})Q^2/(s \, D_1 D_2)$ (right plots) with respect to
$Y_0\equiv\ln(s_0/Q^2)$ and $\mu_R/Q$ for three values of $Y\equiv\ln(s/Q^2)$, at $Q^2$=24 GeV$^2$ and $n_f=5$. These plots were obtained fixing the $\epsilon$ parameter of Eq.~(\ref{VP}) at $10^{-7}$. We have checked the stability of our results under variations of $\epsilon$, verifying that for $Y=10$, when it ranges between $10^{-7}$ and $10^{-4}$, our data points change between $10^{-4}\%$ and $20\%$, the largest variations occurring in a few cases always at points with small absolute values (see Fig~\ref{epsilon}, up). When $Y$ decreases, the region of large $Y_0$ and small $\mu_R$ is affected by a more significant variation, which reaches $46\%$ for $Y=2$ (see Fig~\ref{epsilon}, down). This effect does not compromise the validity of our results, given that it arises far from the stability region of the amplitude in the standard representation. Moreover, for the series representation, to be considered in the next subsection, under the same variation of $\epsilon$ data points change 
between $10^{-6}\%$ and $10^{-2}\%$, leaving the amplitude practically unaffected.

The behavior of the amplitudes in the two basis for the standard representation is very different. In the LO basis, ${\cal A}_0$ becomes stable for high values of $\mu_R$ and $Y_0$, but for the "natural" settings $Y_0=0$ and $\mu_R=Q$ it takes large negative values. In the NLO basis, the amplitude exhibits a global maximum (which becomes local at smaller $Y$) in correspondence of $Y_0=0$ and $\mu_R=4Q$. Again, for higher values of $Y_0$ and $\mu_R$ the amplitude becomes rather stable.
Notice also that the value of ${\cal A}_1$ at the maximum is generally different from the value of ${\cal A}_0$ in the plateau-like region.

\begin{multicols}{2}
    \end{multicols}
\begin{multicols}{2}
    \end{multicols}
\begin{multicols}{2}
    \end{multicols}
    
\subsection{Series representation}
It is also interesting to analyze the behavior of the amplitudes in the two basis using a different representation. Following Ref.~\cite{Ivanov:2005gn}, we recast both amplitudes (\ref{A_0}) and (\ref{A_1}) in the so-called "series representation", {\it i.e.} we rewrite them making explicit the summation over leading and next-to-leading energy logarithms. After some straightforward manipulations, the final form of the series representation will be the same for both ${\cal}A_0$ and ${\cal}A_1$: 
\begin{equation}
\frac{Q^2}{D_1 D_2} \frac{\Im m_s {{\cal A}_{0,1}}}{s} =
\frac{1}{(2\pi)^2}  \alpha_s^2 \label{honest_NLA1} \notag \\
\end{equation}
\begin{equation}
\label{series1}
\times \biggl[ b_0
+\sum_{n=1}^{\infty}\bar \alpha_s^n   \, b_n \,
\biggl(\ln\left(\frac{s}{s_0}\right)^n   +
d_n(s_0,\mu)\ln\left(\frac{s}{s_0}\right)^{n-1}     \biggr)
\biggr]\;,
\end{equation}
where the coefficients of the leading logs, $b_n$, are the same in the two bases,
\begin{equation}
\frac{b_n}{Q^2}=\int\limits^{+\infty}_{-\infty}d\nu \,  c_1(\nu)c_2(\nu)
\frac{\chi^n(\nu)}{n!}\; , \notag
\end{equation}
while the coefficients of the next-to-leading logs,
$d_n$, read as follows, respectively in the LO basis~\cite{Ivanov:2005gn} and in the NLO one:
\begin{equation}
    d_n\bigr|_{\rm LO \ basis}=n\ln\left(\frac{s_0}{Q^2}\right)+\frac{\beta_0}{4N_c}
\Bigg[(n+1) \frac{b_{n-1}}{b_n}\ln\left(\frac{\mu_R^2}{Q^2}\right)
-\frac{n(n-1)}{2}  \notag
\end{equation}
\begin{equation}
    + \frac{Q^2}{b_n}\int\limits^{+\infty}_{-\infty}d\nu \, (n+1)\, f(\nu)
c_1(\nu)c_2(\nu)
\frac{\chi^{n-1}(\nu)}{(n-1)!}\Bigg] \notag
\end{equation}
\begin{equation}
    +\frac{Q^2}{b_n}\left(
\int\limits^{+\infty}_{-\infty}d\nu\, c_1(\nu)c_2(\nu)
\frac{\chi^{n-1}(\nu)}{(n-1)!}\left[
\frac{\bar c^{(1)}_{1}(\nu)}{c_{1}(\nu)}+\frac{\bar
c^{(1)}_{2}(\nu)}{c_{2}(\nu)}
 +(n-1)\frac{\bar \chi(\nu)}{\chi(\nu)}\right]
\right)\;,
\end{equation}
\begin{equation}
    d_n\bigr|_{\rm NLO \ basis}=\frac{Q^2}{b_n}\frac{n}{2}\ln\left(\frac{s_0}{Q^2}\right)\notag
\end{equation}
\begin{equation}
    \times \int\limits^{+\infty}_{-\infty}d\nu \,\Bigg(\frac{1}{1+\bar\alpha_s(A_\nu I_1+B_\nu I_2)}+\frac{1}{1+\bar\alpha_s(A^*_\nu I_1^*+B^*_\nu I^*_2)}\Bigg)c_1(\nu)c_2(\nu)
\frac{\chi^{n}(\nu)}{(n)!} \notag
\end{equation}
\begin{equation}
+\frac{\beta_0}{4N_c}
\Bigg[-\frac{b_{n-1}}{b_n}\ln\left(\frac{\mu_R^2}{Q^2}\right)
-\frac{n(n-1)}{2} +\frac{5}{3}n\frac{b_{n-1}}{b_n}
    + \frac{Q^2}{b_n}\int\limits^{+\infty}_{-\infty}d\nu \, c_1(\nu)c_2(\nu)
\frac{\chi^{n-1}(\nu)}{(n-1)!}
 \notag
\end{equation}
\begin{equation}
    \times\Bigg(\Bigg(f(\nu)+\ln\left(\frac{\mu_R^2}{Q^2}\right)\Bigg)\Bigg(\frac{1}{1+\bar\alpha_s(A_\nu I_1+B_\nu I_2)}+\frac{1}{1+\bar\alpha_s(A^*_\nu I_1^*+B^*_\nu I^*_2)}\Bigg)-n\frac{5}{3} \notag
\end{equation}
\begin{equation}
+(n-1)f(\nu) +n\ln\left(\frac{\mu_R^2}{Q^2}\right)\Bigg)
    +\frac{Q^2}{b_n}\int\limits^{+\infty}_{-\infty}d\nu \frac{\chi^{n}(\nu)}{(n-1)!} c_1(\nu)c_2(\nu)\Bigg(\frac{if_1(\nu)}{2\chi'(\nu)}+
    \frac{if_2(\nu)}{\chi'(\nu)}
\Bigg(-\frac{5}{3}+f(\nu)\Bigg) \notag
\end{equation}
\begin{equation}
   +\frac{if_4(\nu)}{\chi'(\nu)}
\Bigg)\Bigg] +\frac{Q^2}{b_n}\left(
\int\limits^{+\infty}_{-\infty}d\nu\, c_1(\nu)c_2(\nu)
\frac{\chi^{n-1}(\nu)}{(n-1)!}\left[
\frac{\bar c^{(1)}_{1}(\nu)}{\tilde c_{1}(\nu)}+\frac{\bar
c^{(1)}_{2}(\nu)}{\tilde c_{2}(\nu)}
 +(n-1)\frac{\bar \chi(\nu)}{\chi(\nu)}\right]
\right)\;,
\label{dn1}
\end{equation}
with
\begin{equation}
    f_4(\nu)= \psi'(-i\nu+\frac{3}{2}) -2\psi' (-2i\nu+3)- \psi'(i\nu+\frac{3}{2})  + 2\psi' (2i\nu+3)\ .
\end{equation}
To obtain the latter expression we choose to deal with the terms singular at $\nu=0$ in the integrand by an integration by parts, rather than through the principal-value prescription, which we use to regularize just $B_\nu$.
We again notice that some terms beyond the next-to-NLA are included through the use of $\tilde c_{1,2}$ at the denominator. 

\begin{figure*}
            \includegraphics[width=.45\textwidth]{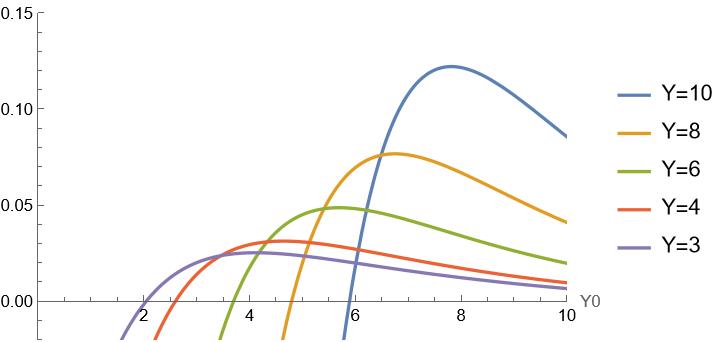}\hfill
            \includegraphics[width=.45\textwidth]{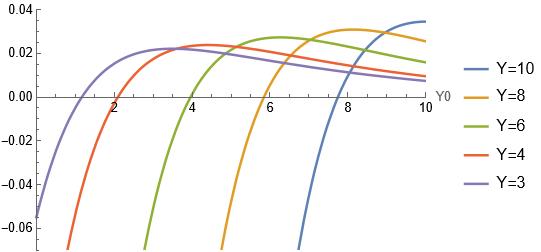}\hfill
            \caption{For $\mu_r=Q$, ${\cal I}m_s ({{\cal A}_0})Q^2/(s \, D_1 D_2)$ on the left and ${\cal I}m_s ({{\cal A}_1})Q^2/(s \, D_1 D_2)$ on the right.}
            \label{muR=Q}
        \end{figure*}

\begin{figure*}
            \includegraphics[width=.45\textwidth]{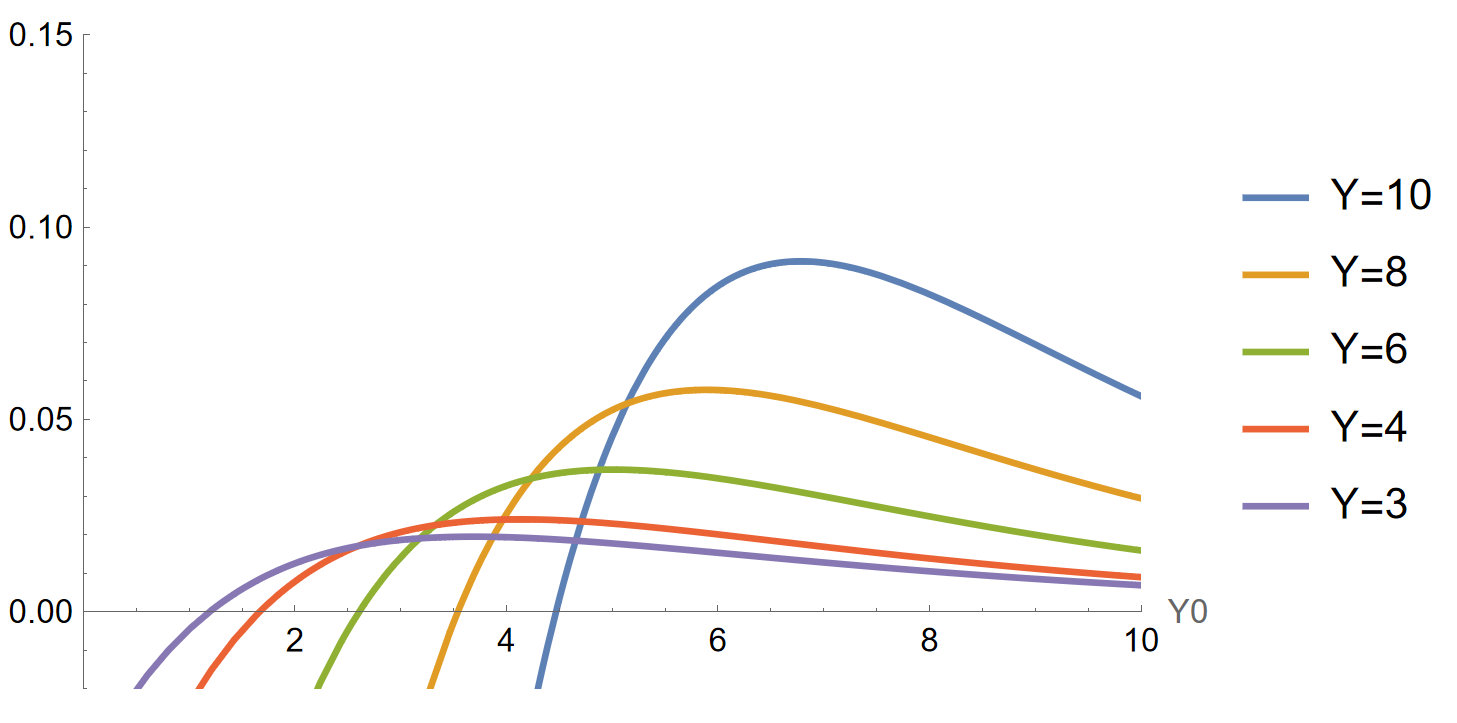}\hfill
            \includegraphics[width=.45\textwidth]{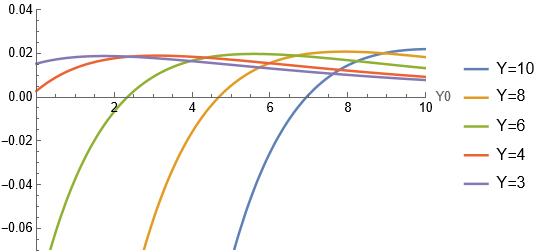}\hfill
            \caption{For $\mu_r=2Q$, ${\cal I}m_s ({{\cal A}_0})Q^2/(s \, D_1 D_2)$ on the left and ${\cal I}m_s ({{\cal A}_1})Q^2/(s \, D_1 D_2)$ on the right.}
            \label{muR=2Q}
        \end{figure*}

\begin{figure*}
            \includegraphics[width=.45\textwidth]{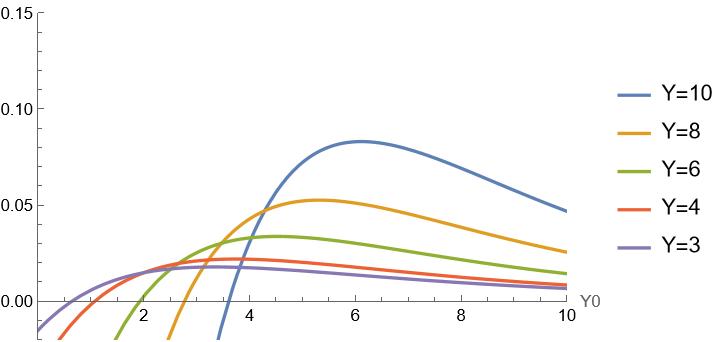}\hfill
            \includegraphics[width=.45\textwidth]{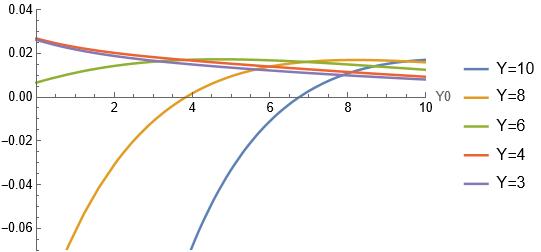}\hfill
            \caption{For $\mu_r=3Q$, ${\cal I}m_s ({{\cal A}_0})Q^2/(s \, D_1 D_2)$ on the left and ${\cal I}m_s ({{\cal A}_1})Q^2/(s \, D_1 D_2)$ on the right.}
            \label{muR=3Q}
        \end{figure*}

In the following numerical analysis, we truncated the series to the first 20 term, as in Ref.~\cite{hep-ph/0508162}. In Figs~\ref{muR=Q}, \ref{muR=2Q}, \ref{muR=3Q} we present, at three fixed different values of $\mu_R$,
the behavior of ${\cal I}m_s ({{\cal A}_0})Q^2/(s \, D_1 D_2)$ (left plots) and ${\cal I}m_s ({{\cal A}_1})Q^2/(s \, D_1 D_2)$ (right plots) with respect to $Y_0$ for several choices of $Y$, at $Q^2$=24 GeV$^2$ and $n_f=5$. The results for the amplitude in the LO basis are identical to those obtained long ago in Ref.~\cite{Ivanov:2005gn}. When the NLO basis is used, the shape of curves is rather different:
i) at smaller values of $\mu_R$ ($\mu_R=Q,2Q$), the amplitude reaches a stationary point (a maximum), while it becomes unstable for $\mu_R=3 Q$ (and at higher values); ii) generally, the value taken at the maximum is smaller in the case of the NLO basis; iii) at smaller values of $Y$, the value of $Y_0$ where the maximum is reached is smaller with the NLO basis; the opposite occurs at the higher values of $Y$.

\begin{multicols}{2}
    \end{multicols}
\begin{multicols}{2}
    \end{multicols}
\begin{multicols}{2}
    \end{multicols}

\section{Conclusions}
\label{sec:conclusions}
We have considered the eigenfunctions of the NLO BFKL kernel in the construction by Chirilli and Kovchegov and used them to get a general expression for the amplitude of forward exclusive hadronic processes in the semihard regime of perturbative QCD.
Similarly to the basis of LO eigenfunctions of the BFKL kernel, the NLO eigenfunctions form a complete set in the transverse momentum space, labeled by a discrete index, $n$, and a continuum real parameter, $\nu$; however, differently from the LO basis, they are singular in $\nu=0$ and must therefore be understood as distributions, subject to a suitable principle-value prescription.

We have first checked that, within the strict next-to-leading logarithmic approximation and in the $n=0$ sector, the expressions of the amplitudes written in the two bases are in perfect agreement, as it must necessarily be. However, this check was non-trivial due to the presence of the singular terms at $\nu=0$ in the NLO basis. We have thus obtained a handy formula for the amplitude, written in terms of the NLO kernel eigenfunctions, ready to use in future phenomenological applications and easily extensible to inclusive backward/forward processes, where also sectors with $n \neq 0$ can play a role.

We have finally presented a first phenomenological application to the forward electroproduction of two light vector mesons. We have shown that, due to the unavoidable presence of terms beyond the NLLA, numerical implementations of the forward amplitude using with the LO and the NLO bases exhibit a different dependence on the scale parameters entering the BFKL approach, namely the renormalization scale $\mu_R$ and the Mellin scale $s_0$. This implies that the region of values of these parameters where the amplitudes are the least dependent on their variation is different for the two bases, leading to different physical predictions. There is a hint that the use of the NLO basis has stability regions at values of $\mu_R$ and $s_0$ closer to the "natural" value dictated by the kinematics of the process, to be further investigated in other processes. In this respect, a useful test-field could be provided by Mueller-Navelet jet production~\cite{Mueller:1986ey} and, in particular, by the azimuthal correlations of the backward and forward identified jets~\footnote{Shortly after the submission of this paper, a nice work appeared~\cite{Chernyshev:2025kfe}, which studied the Mueller-Navelet jet production with the NLO BFKL eigenfunctions.}.

Another possible future direction is the careful treatment of the low-$q^2$ region, corresponding to $\nu=0$, which is sensitive to collinear effects. These have been disregarded in the present work, which was focused on the mere effect of adopting the NLO basis instead of the LO one. However, the collinear improvement of the BFKL approach \cite{Salam:1998tj,Ciafaloni:1999yw,Ciafaloni:2003rd,SabioVera:2005tiv} is a necessary step towards precision predictions.

\section{Acknowledgments}

The work has been partially funded by the European Union – Next Generation EU through the research grant number P2022Z4P4B “SOPHYA - Sustainable Optimised PHYsics Algorithms: fundamental physics to build an advanced society” under the program PRIN 2022 PNRR of the Italian Ministero dell’Universit\`a e Ricerca (MUR).
The work of M.F. is supported by the ULAM fellowship program of NAWA No. BNI/ULM/2024/1/00065 ``Color glass condensate effective theory beyond the eikonal approximation''. The work of A.Pa. is supported by the INFN/QFT@COLLIDERS project (Italy).

\appendix

\section{Eigenvalues of the NLO BFKL kernel in the azimuthally-dependent case}

In this Appendix we show the proof of~(\ref{eigen_gen_n}). We first introduce for the NLO eigenstates of the BFKL kernel the notation 
\[
|\nu,n\rangle = |\nu^{(0)},n\rangle + \bar\alpha_s |\nu^{(1)},n\rangle \;,
\]
where
\[
\langle \vec q|\nu^{(0)},n\rangle = \phi_{\nu}^{(0)}(\vec q^{\:2}) e^{i n \theta}
\equiv \phi_{\nu,n}^{(0)}(\vec q)\;,
\]
\[
\langle \vec q|\nu^{(1)},n\rangle = \phi_{\nu,n}^{(0)}(\vec q) \ h_{\nu,n}(\vec q^{\:2})\;,
\]
with $\phi_\nu^{(0)}(\vec q^{\:2})$ and $h_{\nu,n}(\vec q^{\:2})$ defined, respectively, in~(\ref{LO_eigen}) and~(\ref{eigenf_nu_n_final}).
Then, recalling~(\ref{K_expansion}), we have that, within the NLO, 
\[
\hat K \, |\nu,n \rangle \simeq \bigl(\bar\alpha_s \hat K^{(0)} 
+ \bar\alpha_s^2 \hat K^{(1)}\bigr) |\nu^{(0)},n\rangle
+ \bar\alpha_s^2 \hat K^{(0)} |\nu^{(1)},n\rangle \;.
\]
From Refs.~\cite{Kotikov:2000pm,Kotikov:2002ab}, we know that
\beq
\bigl(\bar\alpha_s \hat K^{(0)} + \bar\alpha_s^2 \hat K^{(1)}\bigr) 
|\nu^{(0)},n\rangle = \bar \alpha_s \chi(\nu,n) |\nu^{(0)},n\rangle
\eeq
\[
+ \bar\alpha_s^2
\left[\chi_1(\nu,n)+\frac{\beta_0}{4N_c}\left(\chi(\nu,n) \ln \mu_R^2+\frac{i}{2}\chi'(\nu,n)+i\chi(\nu,n)\frac{\partial}{\partial\nu}\right)\right]
|\nu^{(0)},n\rangle \;,
\]
which implies
\beq
\langle \vec q | \bigl(\bar\alpha_s \hat K^{(0)} + \bar\alpha_s^2 \hat K^{(1)}\bigr)  |\nu^{(0)},n\rangle 
= \bar \alpha_s \chi(\nu,n) \phi_{\nu,n}^{(0)}(\vec q) 
\label{piece_1}
\eeq
\[
+ \bar\alpha_s^2
\left[\chi_1(\nu,n)+\frac{\beta_0}{4N_c}\left(\chi(\nu,n) \ln \frac{\mu_R^2}{\vec q^{\:2}}+\frac{i}{2}\chi'(\nu,n)\right)\right] \phi_{\nu,n}^{(0)}(\vec q) \;.
\]

On the other side, 
\[
\langle \vec q | \hat K^{(0)} |\nu^{(1)},n\rangle 
= \int d^{D-2}k \, K^{(0)}(\vec q, -\vec k) \phi_{\nu,n}^{(0)}(\vec k) 
\, h_\nu(|\vec k|) \;,
\]
with $D=4+2\epsilon$ to regularize infrared divergences which will appear in intermediate steps and 
\[
\bar \alpha_s K^{(0)}(\vec q, -\vec k) = \underbrace{2 \omega^{(1)}(\vec q^{\: 2}) \, \delta^{(D-2)}(\vec q - \vec k)}_{\rm virtual \ part} + \underbrace{\frac{g^2 N_c}{(2\pi)^{D-1}}\frac{2}{(\vec q - \vec k)^2}}_{\rm real \ part}\;,
\]
where
\[
\omega^{(1)}(\vec q^{\: 2}) = -\frac{g^2 N_c \Gamma(1-\epsilon)}{(4\pi)^{2+\epsilon}} \frac{\Gamma^2(\epsilon)}{\Gamma(2\epsilon)} (\vec q^{\: 2})^\epsilon
\]
is the LO gluon Regge trajectory. We recall that, in dimensional regularization, $g$ has mass dimension equal to $-\epsilon$ and $g^2$ is meant everywhere as $\tilde g^2 \mu^{-2 \epsilon}$, with $\tilde g$ the dimensionless coupling and $\mu$ the renormalization scale deemed to finally become $\mu_R$.
The contribution from the virtual part of the LO kernel is trivial:
\[
\bar\alpha_s \langle \vec q | \hat K^{(0)} |\nu^{(1)},n\rangle \Bigr|_{\rm virt.}
= -\frac{2 g^2 N_c \Gamma(1-\epsilon)}{(4\pi)^{2+\epsilon}} \frac{\Gamma^2(\epsilon)}{\Gamma(2\epsilon)} \frac{(\vec q^{\: 2})^{i\nu-1/2+\epsilon}}{\pi\sqrt{2}} e^{i n \theta}
\left[A_{\nu,n}\ln^2\frac{\vec q^{\:2}}{\mu_R^2}
+B_{\nu,n}\ln\frac{\vec q^{\:2}}{\mu_R^2}\right]\;.
\]
The contribution from the real part of the LO kernel is given by the
following integral:
\[
\bar\alpha_s \langle \vec q | \hat K^{(0)} |\nu^{(1)},n\rangle \Bigr|_{\rm real}
= \frac{g^2 N_c}{(2\pi)^{D-1}} \!\int\! d^{D-2}k 
\frac{2}{(\vec q - \vec k)^2} \frac{(\vec k^{\,2})^{i\nu-1/2}}{\pi\sqrt{2}}
e^{i n \phi} \!\left[A_{\nu,n}\ln^2\frac{\vec k^{\,2}}{\mu_R^2}
+B_{\nu,n}\ln\frac{\vec k^{\,2}}{\mu_R^2}\right],
\]
where $\phi$ is the azimuthal angle of $\vec k$. To calculate this integral, 
it is convenient to use the representation
\[
e^{in\phi} = \frac{(\vec k \cdot l)^n}{(\vec k^{\,2})^{n/2}}\;,
\]
where $\vec l$ lies in the first two of the $D-2=2+2\epsilon$ transverse space dimensions, {\it i.e.} $\vec l = \vec e_1 + i \vec e_2$, with $\vec e_{1,2}^{\:2}=1$
and $\vec e_1 \cdot \vec e_2=0$. The second trick is to write
\[
\ln \vec k^{\,2} = \frac{\partial}{\partial a}\biggr|_{a\to 0} (\vec k^{\,2})^a\;,
\;\;\;\;\;
\ln^2 \vec k^{\,2} = \frac{\partial^2}{\partial a^2}\biggr|_{a\to 0} (\vec k^{\,2})^a\;,
\]
and to take derivatives out the integral. Then, the Feynman parametrization can be used,
\[
\frac{1}{A_1^{\alpha_1} A_2^{\alpha_2}} = \frac{\Gamma(\alpha_1+\alpha_2)}{\Gamma(\alpha_1)\Gamma(\alpha_2)}\, \int_0^1 dx\, \frac{x^{\alpha_1-1} (1-x)^{\alpha_2-1}}{[x A_1 + (1-x) A_2]^{\alpha_1+\alpha_2}}\;,
\]
with $A_1=(\vec q - \vec k)^2$, $\alpha_1=1$ and $A_2=\vec k^{\,2}$, $\alpha_2=(n+1)/2-i\nu-a$, to completely solve the integral and get, after some simple manipulations, the following result (see Sect.~A.3 of Ref.~\cite{Fucilla:2023pma} for details about the calculation of this kind of integrals):
\[
\bar\alpha_s \langle \vec q | \hat K^{(0)} |\nu^{(1)},n\rangle \Bigr|_{\rm real}
= \frac{4g^2 N_c \Gamma(\epsilon)}{(4\pi)^{2+\epsilon}} 
\left\{A_{\nu,n} \frac{\partial^2}{\partial a^2}\biggr|_{a\to 0}
+ B_{\nu,n} \frac{\partial}{\partial a}\biggr|_{a\to 0}\right\}
\]
\[
\times
\left[\frac{\Gamma(i\nu+1/2+n/2+a+\epsilon)\, \Gamma(-i\nu+1/2+n/2-a-\epsilon)}
{\Gamma(i\nu+1/2+n/2+a+2\epsilon)\, \Gamma(-i\nu+1/2+n/2-a)}\,
\frac{(\vec q^{\:2})^{i\nu-1/2+a+\epsilon}}{\pi\sqrt{2}}\right] e^{i n \theta}\;.
\]
It can be easily checked that, in the sum of the real and virtual parts the pole in $\epsilon$ disappears, so that $\epsilon$ can be safely put equal to zero, to get
\[
\bar\alpha_s \langle \vec q | \hat K^{(0)} |\nu^{(1)},n\rangle = \bar \alpha_s \, 
\phi_{\nu,n}^{(0)}(\vec q)\, 
\left[\chi(\nu,n)\left(A_{\nu,n} \ln^2\frac{\vec q^{\:2}}{\mu_R^2} 
+ B_{\nu,n}\ln\frac{\vec q^{\:2}}{\mu_R^2}\right) \right.
\]
\beq
\left. + \frac{\beta_0}{4N_c} \left(\chi(\nu,n) \ln\frac{\vec q^{\:2}}{\mu_R^2} -i\frac{\chi'(\nu,n)}{2}\right)
\right]
\label{piece_2}
\eeq
\[
= \bar\alpha_s \chi(\nu, n)\, \phi_{\nu,n}^{(0)}(\vec q) h_{\nu,n}(\vec q^{\:2}) + \bar\alpha_s \phi_{\nu,n}^{(0)}(\vec q) \frac{\beta_0}{4N_c} \left(\chi(\nu,n) \ln\frac{\vec q^{\:2}}{\mu_R^2} -i\frac{\chi'(\nu,n)}{2}\right)
\;.
\]
Finally, putting together the results in~(\ref{piece_1}) and~(\ref{piece_2}), one gets, within NLO accuracy,
\[
\langle \vec q | \hat K | \nu,n\rangle \simeq
\langle \vec q | \bigl(\bar\alpha_s \hat K^{(0)} + \bar\alpha_s^2 \hat K^{(1)}\bigr)  |\nu^{(0)},n\rangle + \langle \vec q | \bar\alpha_s^2  \hat K^{(0)} |\nu^{(1)},n\rangle 
\]
\[
= \bar \alpha_s \chi(\nu,n) \,\phi^{(0)}_{\nu,n}(\vec q^{\:2})\biggl(1+\bar\alpha_s h_{\nu,n}(\vec q^{\:2})\biggr) + \bar\alpha_s^2 \chi_1(\nu,n) \phi^{(0)}_{\nu,n}(\vec q^{\:2}) 
\]
\[
\simeq \Delta(\nu,n) \langle \vec q| \biggl(|\nu^{(0)},n\rangle +\bar \alpha_s |\nu^{(1)},n\rangle \biggr) = \Delta(\nu,n) \langle \, \vec q| \nu,n\rangle \;,
\]
with 
\[
\Delta(\nu,n)= \bar \alpha_s(\mu_R) \chi(\nu,n) + \bar\alpha_s^2(\mu_R)
\chi_1(\nu,n) \;.
\]

\newpage
\bibliographystyle{apsrev}
\bibliography{references}

\end{document}